\documentclass[fleqn,usenatbib]{mnras}
\pdfoutput=1
\usepackage{newtxtext,newtxmath}

\usepackage[T1]{fontenc}
\usepackage{ae,aecompl}

\usepackage{graphicx} 
\usepackage{dcolumn} 
\usepackage{bm} 
\usepackage{amsfonts}
\usepackage{amsmath}
\usepackage{amssymb}
\usepackage{epsfig}

\usepackage{color}
\usepackage{hyperref}
\usepackage[utf8]{inputenc}
\usepackage[english]{babel}
\usepackage{enumitem}
\usepackage{breakcites}

\newcommand{\ntot}{N_\text{tot}}
\newcommand{\mtot}{M_\text{tot}}
\newcommand{\ndet}{N_\text{det}}

\newcommand{\pmin}{p_\text{min}}
\newcommand{\nbio}{N_\text{bio}}
\newcommand{\fbio}{f_\text{bio}}
\newcommand{\fabio}{f_\text{abio}}
\newcommand{\rdet}{r_\text{det}}
\newcommand{\ntop}{N_\text{top}}
\newcommand{\mum}{\mu\text{m}}
\newcommand{\methane}{\text{CH}_4}

\newcommand{\oxygen}{\text{O}_2}
\newcommand{\ozone}{\text{O}_3}
\newcommand{\Odimer}{\text{O}_4}
\newcommand{\nitrous}{\text{N}_2\text{O}}
\newcommand{\nitrogen}{\text{N}_2}
\newcommand{\carbon}{\text{CO}_2}

\newcommand{\beq}{\begin{equation}}
\newcommand{\eeq}{\end{equation}}
\newcommand{\beqa}{\begin{eqnarray}}
\newcommand{\eeqa}{\end{eqnarray}}

\title[Biosignature Surveys]{Biosignature Surveys to Exoplanet Yields and Beyond}

\author[Sandora and Silk]{
	McCullen Sandora,$^{1,2}$\thanks{E-mail: mccullen.sandora@gmail.com}
	Joseph Silk,$^{2,3,4}$\thanks{E-mail: joseph.silk@physics.ox.ac.uk}
	\\
	$^{1}$Center for Particle Cosmology, University of Pennsylvania, Philadelphia, PA 19104, USA\\
	$^{2}$Institut d’Astrophysique de Paris, 98 bis Bd. Arago, F-75014 Paris, France\\
	$^{3}$Department of Physics and Astronomy, Johns Hopkins University, Baltimore, MD 21218, USA\\
	$^{4}$BIPAC, Department of Physics, University of Oxford, Keble Road, Oxford OX1 3RH, UK
}

\pubyear{2020}

\begin{document}
	\label{firstpage}
	\pagerange{\pageref{firstpage}--\pageref{lastpage}}
	\maketitle
	
	\begin{abstract}
		Upcoming biosignature searches focus on indirect indicators to infer the presence of life on other worlds. Aside from just signaling the presence of life, however, some biosignatures can contain information about the state that a planet's biosphere has achieved. This additional information can be used to measure what fractions of planets achieve certain key stages, corresponding to the advent of life, photosynthesis, multicellularity, and technological civilization. We forecast the uncertainties of each measurement for upcoming surveys, and outline the key factors that determine these uncertainties. Our approach is probabilistic and relies on large numbers of candidates rather than detailed examination of individual exoplanet spectra. The dependence on survey size, likeliness of the transition, and several measures of degrees of confidence are discussed, including discussion of geological false positives in biosignatures as well as how combining data from different missions can affect the inference. Our analysis should influence policy recommendations for future mission design and strategy to minimize the impact of measurement uncertainties.
		
	\end{abstract}
	
	
	
	
	\section{Introduction}
	
	The era of large exo-atmosphere surveys will quickly be upon us. Near future experiments aim to measure the atmospheres of several Earthlike planets, and in a few decades, large scale surveys will collect data on dozens of Earthlike worlds (\cite{kiang2018exoplanet}). This is exciting because life can potentially have a large impact on a planet's atmosphere, and so the measurement of the gas content of an exoplanet atmosphere can serve as an indirect detection of the presence of any life (\cite{lovelock1965physical}). We address the following issue via  a probabilistic framework: what is the optimal number of exoplanet targets? Our arguments are simple: we do not present detailed spectral analyses of exoplanet atmospheres, undoubtedly a crucial ingredient, but we focus on giving reasonable odds for success based on having a sufficient number of targets and making plausible assumptions about the nature of biosignatures.
	
	The stated goal of many missions has been to clearly detect atmospheric gases in as many systems as possible. 
	However, biosignatures may carry more information than just the presence or absence of life: it is also possible to deduce the level of sophistication that life on a planet has attained, relying on the fact that different signatures signal alternative biochemical processes. 
	The life history of our own planet can be seen as a sequence of transitions wrought by evolutionary innovations, from biogenesis to the evolution of photosynthesis, multicellularity, and technological civilization (\cite{majortrans}). As far as these transitions can be expected to be generic, they can each be sought for independently through their characteristic atmospheric imprints. The question we address here is, what fraction of planets undergoes each transition, and more importantly, which can be measured with upcoming surveys?  By quantifying the uncertainty in measurements of each of these quantities, we provide a framework for understanding how they depend on proposed mission designs as well as on atmospheric modeling. If the goal is to maximize the information return to the extent possible, this also provides a means of policy recommendation for which aspects of future missions most effort should be devoted.
	
	We begin in section \ref{sign} with an overview of each of the different transitions that we think are important, and some ideas for the associated signatures that can be measured. Not all of these are likely to be constrained in the immediate future, but enunciating our ultimate desires for science return can then serve to focus future research into some of the more challenging measurements.  In section \ref{tele} we review current and future telescope missions, and their expected return on each of these measurements. Then, we outline the formalism for parameter estimation, as a function of number of surveyed planets and detections: we begin in section \ref{sing} with the simple case of detecting a single signal, and outline how this is affected by priors. In section \ref{mult}, we discuss the detection of multiple signals in a single data set, and find that most effort should be spent measuring signals that correspond to likely transitions. Section \ref{comb} is devoted to the subtleties that arise when combining multiple, possibly incomplete datasets, and we find recommendations for which mission yields to optimize. In section \ref{conf} we outline how false positives and negatives influence the variance of our likelihoods, and find that generically the desired confidence in our signal interpretation should be roughly the same magnitude as the likelihood of the signal. Finally, in section \ref{disc}, we provide a general prescription for maximizing science return.
	
	\section{Signatures}\label{sign}
	
	We focus here on what we expect to be a nested sequence of biosignatures, corresponding to the major evolutionary transitions that occur on Earth, as outlined in \cite{carterbio,majortrans}. Many of the biosignatures 
	reorganize the energy flow of the entire biosphere, leaving a large imprint on the planet, most notably via  the vast increase in available energy that accompanies these transitions.
	
	The first step we take will be the presence of life in any form. The next step is photosynthesis, the harvesting of energy from the planet's host star thereby
	providing
	a mechanism to harvest the dominant source of free energy on a planet. After this, we discuss the evolution of multicellularity on other worlds. Lastly, and most speculatively, the detection of technological civilizations, both of the level Earth has achieved and beyond, is considered. 
	Though most work has been done on detecting photosynthesis and technosignatures, and these are likely the only ones which will be unambiguously measurable with the next generation of telescopes, our goal is to call attention to the importance of distinguishing the other two, with the hopes of spurring the creativity needed to make measuring these a reality.
	
	\subsection{Life}
	
	Any form of life is expected to be accompanied by chemosynthesis, the rearrangement of chemical matter. 
	Several simple gaseous byproducts have been identified that are indicative of life. It is important to note that most of the biosignatures that we discuss do not unambiguously signal the presence of life, however: 
	several abiotic sources of these gases have been identified that could give rise to false positives. Identifying combinations of observables that would strengthen our confidence that life is the only possible source (type III biosignatures in the terminology of \cite{seager2013biosignature}) is an important industry, but here we wish to merely include a sampling of the types of signatures that have been proposed.  Similarly, detecting the presence of one of these signatures will not guarantee that none of the other, later stages have not been achieved, only that this one has.
	
	The most studied byproduct of simple is is methane ($\methane$) (\cite{guzman2013abiotic}).	Methanogenesis evolved very early on Earth (\cite{battistuzzi2004genomic}), and
	is best detected in the 1.0-1.7 $\mum$ wavelength regime. Several of its spectral peaks coincide with those of water, but observing the 0.84 and 0.92 $\mum$ features at SNR$>5$ will unambiguously signal detection (\cite{orange}). Additionally, it has a feature at 7.7 $\mum$ that would be suitable for measuring with eclipse spectroscopy on planets whose blackbody spectrum peaks near this location (\cite{fujii_biosig}).
	
	Though the methane on Earth is overwhelmingly produced by life (\cite{etiope2013abiotic}), there are several abiotic production mechanisms, such as the serpentinization of rocks, so that it may not automatically be considered a biosignature, especially if the atmosphere is reducing, as on Titan. Ways to disambiguate these two sources are to look for chemical disequilibria: for example, the simultaneous presence of $\methane$, $\carbon$, $\nitrogen$, and liquid water would signal some prodigious methane source, most easily explained by biology (\cite{krissansen2018disequilibrium}). 
	
	Methane was probably not detectable throughout all of Earth's history; oxygenation of the Proterozoic caused methane levels to drop to undetectable levels (\cite{reinhard2016earth}). It is likely that methane can only be detected when primary productivity is very high, which requires an energy source as ubiquitous as oxygenic photosynthesis (\cite{WRphoto}). However, its associated byproduct, haze, can serve as an important indirect indicator of its presence (\cite{orange}). Haze is only formed (around sunlike stars) when biological levels of methane are present, and can be detected fairly easily from its continuum NUV signature,  most relevant for the Archean eon, when methane levels were 2-3 orders of magnitude higher than present (\cite{pavlov2000greenhouse}). 
	
	It was found in \cite{wang2018baseline} that methane is detectable with HabEx after 850 hours of exposure time with favorable cloud conditions, and after 25 hours with LUVOIR under similar conditions.
	
	An additional chemical signature of life is nitrous oxide $\nitrous$ (\cite{des2002remote}). This gas is produced when other biogenic nitrogen compounds react with atmospheric oxygen, and so requires the presence of oxygenic photosynthesis, in the absence of any other oxygen sources.  In contrast with $\methane$, this biosignature was enhanced during the mid-Proterozoic as a consequence of enhanced oxygen levels (\cite{buick2007did}) and can be observed from its several peaks in the 1.5-2.0 $\mum$ range. Abiotic sources that would need to be taken into account include lightning (\cite{harman2018abiotic}) and flares (\cite{airapetian2016prebiotic}). 
	Additional biosignature gases have been proposed as well, such as methyl chloride (CH$_3$Cl) and dimethyl sulfide (C$_2$H$_6$S) (\cite{seager2016toward}), as well as the generic strategy of searching for chemical disequilibrium. 
	Another generic feature of life is biological homochirality, the preference for one molecular handedness over the other, 
	which we discuss in the next subsection. 
	

	\subsection{Photosynthesis}
	
	The advent of oxygenic photosynthesis, the capability of using sunlight to rip electrons off water molecules, was one of the key innovations in the history of life on Earth. 
	It likely only appeared on Earth sometime between 3.7 Ga (\cite{rosing2004u}) and 2.4 Ga (\cite{kirschvink2008palaeoproterozoic}- and see \cite{lyons2014rise} for a review), though the date of its appearance is somewhat controversial. This innovation had major consequences not only for the size of the Earth's biosphere and types of possible organisms, but also for the entire chemistry of the planet itself: it was a key contributor to the oxygenation of the atmosphere.
	
	Molecular oxygen ($\oxygen$) is best detected by a strong spectral line at 0.76 $\mum$, observation of which will necessitate having a spectral resolution of greater than 70 (\cite{habex}). Detection would require near-present day levels, and if Proterozoic levels were $0.1\%$ of present atmospheric level as some reconstructions indicate, $\oxygen$ would not have been directly observable during this time (\cite{planavsky2014low}).
	
	However, it is much easier to infer the presence of $\oxygen$ indirectly through its photolytic byproduct ozone ($\ozone$). This has many features in the range 0.3-0.7 $\mum$, and on modern Earth manifests itself as a sharp cutoff at 0.33 $\mum$ (\cite{habex}). Ozone would have been readily observable during the Proterozoic, independently of atmospheric oxygen level (\cite{luvoir}). Further, the production of NO and OH can be used to infer the presence of an oxygen atmosphere by their signatures at 5.3 $\mum$ and 1.7 $\mum$, respectively (\cite{airapetian2017atmospheric}).
	
	Several abiotic sources of oxygen have been detailed recently, among them the photodissociation of water and subsequent loss of hydrogen to space (\cite{luger2015extreme,airapetian2017hospitable}). Most known abiotic production mechanisms produce a very high atmospheric O/H ratio, which leads to a lack of water vapor (\cite{gao2015stability}) and clouds (\cite{wordsworth2014abiotic}), and so can be ruled out if atmospheric water content is accurately measured. However, there may be certain planets that flaunt this rule, and so a full contextual analysis must be undertaken if oxygen is detected on any exoplanet (\cite{meadows2018exoplanet}). Determining the water abundance requires observing multiple spectral features throughout the 0.8-2 $\mum$ range, which will be achievable with a spectral resolution of 70 (\cite{luvoir}). Many of the abiotic production mechanisms also lead to a much higher concentration of oxygen than observed on Earth, which would lead to $\Odimer$ absorption features in the 0.3-0.8 and 1.0-1.4 $\mum$ bands that can be searched for (\cite{luger2015extreme}). LUVOIR aims to take advantage of these signals to be able to rule out all known oxygen false positives. Additionally, seasonal $\ozone$ differences may be sought, but these will only occur for low $\oxygen$ levels, where the ozone layer has not saturated to its maximal value (\cite{mason2008exoplanets}).
	
	An additional signature of photosynthesis exploits the fact that chlorophyll and other known photosynthetic pigments have optimized their frequency response profile to maximize the number of photons collected, while simultaneously screening out higher energies that would cause overheating (\cite{ford2001characterization}). This leads to what is known as the red edge, a sharp drop-off in the absorption properties at 0.7 $\mum$, giving rise to the characteristic green appearance of our planet, which is spectrally detectable from space (\cite{arnold2002test}). This was proposed as a promising biosignature for exoplanets in \cite{seager2005vegetation}, and has no known false positives in the wavelength region where we may observe it. It is important to note, however, that since photosynthesis is optimized for the incident light spectrum, the edge may occur in a different part of the spectrum for planets orbiting different mass stars (\cite{WRphoto}). If so, then the reflectance profiles of certain minerals, namely cinnabar and sulfur, which have edges at 0.6 and 0.45 $\mum$, can mimic the edge expected from biology (\cite{schwieterman2018exoplanet}). Being a broadband optical feature, high resolution is not required for detection of this.
	
	Homochirality manifests itself as a $0.01\%$ circular polarization near the red edge (\cite{sparks2009detection}). Remote detection of polarization from light scattered off vegetation has recently been demonstrated in both linear (\cite{berdyugina2016remote}) and circular (\cite{patty2019circular}) configurations, demonstrating the feasibility of this search strategy. LUVOIR will contain a spectropolarimeter, but only in the range of 0.1-0.4 $\mum$. However, it was suggested in \cite{kiang1,kiang2} that the location of the spectral edge is dictated by the stellar spectrum, so that around low mass stars ($\sim 0.3M_\odot$) photosynthesis may be optimized in this range. Polarization data is a core science goal of the ELF mission. A final biosignature to note is biofluorescence (\cite{o2016biofluorescent}), which would signal the downregulation of harmful light to lower frequencies, and may present itself as a detectable afterglow accompanying flares.
	
	\subsection{Multicellularity}
	
	Multicellularity 
	represents an enormous reorganization of the biosphere, and is easily argued to be a prerequisite for intelligent life. 
	All known multicellular organisms are eukaryotes, and it has been argued that eukaryogenesis is a necessary precondition for multicellularity to occur (\cite{lane2010energetics}). As opposed to prokaryotes, eukaryotes have a tightly controlled internal structure which is capable of selectively expressing genes when certain conditions are met (\cite{bains2016likely}). This, more than anything else, is what enabled the multicellular cooperation necessary for microscopic organisms to reorganize into a macroscopic creature with a single germline. Important as eukaryogenesis was for life on Earth, however, it is relatively invisible in terms of remote detection, with no proposed strategies that the authors are aware of. Given the almost immediate corollary of multicellularity, however, this latter transition can serve as a proxy for the former, if we can find ways to search it out.
	
	Many environments on Earth are only capable of being inhabited by extremophiles, which are uniformly unicellular (\cite{grant1998extremophiles}). Additionally, it was recently pointed out that the habitable zone for unicellular life is expected to be much broader than the habitable zone for complex life (\cite{schwieterman2019limited}), the latter being only $20-28\%$ the width of the former by their estimates. 
	Indeed, the complexity that we witness today only began in the Cambrian era 540 Mya, so that 
	the majority of Earth's history was exclusively populated by simple organisms (\cite{marshall2006explaining}).
	
	Detection of multicellularity is much more difficult than the previous steps, and consequently not as much attention has yet been paid to this. One possible avenue of inference is the detection of large forests: on Earth, the majority of land plant biomass is in the form of multicellular organisms, as opposed to the mostly unicellular sea plants (\cite{SHnpp}). This is argued to be a generic requirement of water and mineral collection, which are not as available on land as in the ocean (\cite{niklas1997evolutionary}). 
	The phase dependence of broadband properties of scattered light \cite{fujii2010colors} would provide a method of detecting land forests, as would be multi-pixel imaging of target planets. Several proposed mission designs aim to do just this: ELF can infer sub-continent resolution of several nearby systems in the near future by inverting the time resolved photometry (\cite{elf}), and further afield, hypertelescopes would be able to obtain 30$\times$30 pixel pictures of Earthlike planets at or beyond 3 pc (\cite{hypertelescope}).
	
	However, the presence of widespread lichens may confound efforts to distinguish multicellular from unicellular life.  Indeed, evidence from weathering rates testifies for widespread land life as early as 2.8 Ga on Earth (\cite{stueken2012contributions}) and that it covered a significant portion of the Earth's surface at least as far back as 1.1 Ga (\cite{kenny2001stable}).  Work has been done on detecting lichens remotely via satellite imagery in the visible to mid-infrared bands (\cite{gilichinsky2011mapping}) and, while subdominant on Earth, lichens may potentially grow to cover the majority of a planet's land surface in the absence of multicellular competitors.
	
	A method to distinguish tall forests from surface lichens was developed in \cite{doughty2010detecting}: by measuring the dependence of reflectance on phase angle, the presence of tall shadows can be inferred, even when averaged over the entire planet.  These methods have further been developed and applied to data collected from Earth by the Galileo space probe (\cite{doughty2016detecting}) and with the POLDER satellite (\cite{doughty2020distinguishing}).
	
	It may be possible to infer that a photosynthetic signal is land-based spectrally: for this, it is important to bear in mind that in the ocean, plankton occur throughout the euphotic zone, where light can penetrate. In fact, the chlorophyll maximum occurs 80 meters below the ocean surface (\cite{SHnpp}). This shifts the spectrum of light collected by these cells, and their chlorophyll pigments have evolved to harvest lower frequencies because of this (\cite{kiang2}). While this avenue would be fraught with several stages of indirect inference, it represents a possible method to determine if an observed signal may possibly imply multicellularity.

	\subsection{Technosignatures}
	
	Ultimately, we would like to determine how many other planets host civilizations like our own. There has been ongoing effort to detect evidence of radio communication through SETI (\cite{tarter2001search}), and there is currently a push to search for more general signatures of technological civilization (\cite{technosignatures}). In fact, it may be possible to find traces of technosignatures in exoplanet spectra. In \cite{lin2014detecting}, the possibility of detecting the  industrial pollutants CF${}_4$ and CCl${}_3$F were outlined, where they found it would be possible around white dwarfs with JWST for 1-2 day exposure time for 10x our current terrestrial levels. These pollutants, along with all other chlorofluorocarbons, have no known nontechnological source (\cite{seager2016toward}), and have residence times of 50,000 yr. They can be observed in the 7.8 and 11.6-12.0 $\mum$ bands, respectively.
	
	An additional technosignature includes a solar power panel analog to the red edge found in plants that can indicate stellar energy harvesting (\cite{lingam2017natural}). 
	
	With multipixel resolution,  it can become feasible to search for signatures of metropolises on exoplanets, either through their waste heat or artificial light (\cite{elf}). 
	
	There is no reason to restrict our biosignature searches to developments that the Earth has so far attained.
	If civilization's heat output rivals that of the host  planet, this defines a Kardashev type I civilization (\cite{mullan2018population}). 
	\cite{kuhn2015global} find that planetary infrared anomalies can be detected with a contrast two orders of magnitude greater than could be detected in the visible spectrum.
	
	Similarly, it is possible to search for Kardashev type II and III civilizations by looking for anomalous luminosities of stars and galaxies, respectively. Stars were searched in RAVE and GAIA data in \cite{zackrisson2018seti}, but of the some 8,000 suitable stellar targets, no clear detection occurred. Galactic sized civilizations were searched for in WISE data in \cite{wright2014g2}, but again, none were detected.

	\section{Telescopes}\label{tele}

	It is important to determine the total number of systems that can be observed with a given technology, in order to determine the mission parameters that will maximize the scientific return. This has been treated in many places (\cite{agol2007rounding,stark2014maximizing,stark2016maximized}). Here, we provide a simple analytic approximation of the total exoplanet yield, which will facilitate comparison between the different missions we consider. For definiteness, we define the yield as the number of Earthlike planets around main sequence stars. 
	
	The time it takes a telescope to make a measurement with a signal-to-noise ratio of 10 can be summarized from \cite{stark2014maximizing} as
	\beq
	\tau=10^3\,\frac{(24\,r_p+4.2\,\zeta)D_t^2+431\,r_z\,\lambda^2}{\phi_\star\, r_p^2\,D_t^4\,\Delta\lambda}\label{tau}
	\eeq
	Here $r_p$ is the ratio of planet to star flux, $\zeta$ is the starlight suppression factor, $r_z$ is the ratio of zodiacal light to star flux, $\phi_\star$ is the stellar photon flux per unit wavelength, $D_t$ is the telescope diameter, and $\Delta\lambda$ is the width of the passband. The first term in the numerator is the noise arising from photon count number, and the second from the background of the target stars. The third term represents a combined contribution of solar and extrasolar zodiacal light, though this depends on assumptions on its prevalence around other star systems. This will become better measured with future missions (\cite{weinberger2015target}). The numerical coefficients will depend on design efficiencies and the particular signal measured, but what will be more important to us are the scalings with the parameters. 
	
	We recap the various mission designs, and how they affect the required integration time: first, upcoming telescopes are planned that are both space-based and ground-based. Ground-based telescopes can be larger, but the Earth's atmosphere sets a lower limit on the contrast that can be achieved at $\zeta=10^{-8}$ (\cite{wang2018baseline}). There, the authors find that the second term in eqn (\ref{tau}) dominates the first unless $\zeta<10^{-8}$ for M dwarfs and $10^{-10}$ for sunlike stars (\cite{wang2018baseline}), and so red dwarfs will be the only suitable targets for upcoming ground-based experiments. 
	
	Telescopes can detect exoplanets through the transit method or direct imaging. 
	Most missions we discuss use a coronagraph, 
	or a starshade.
	The main advantage of the latter is a significantly decreased inner working angle. Starshade missions will also face constraints from fuel and repositioning time (\cite{stark2016maximized}), but this is not taken into consideration in our analysis.
	
	The first two terms in eqn (\ref{tau}) scale with the distance to the target $d$ as $d^2$. If these are dominant, then we can compute the number of systems observed if a telescope operates for a total time $T$. To do so, we use the continuum limit, where observable systems are uniformly distributed throughout space with density $n$, equal to the fraction of stars that possess Earthlike planets multiplied by the density of stars in our local neighborhood (further multiplied by the probability for transit alignment for telescopes using the transit method). The total number of signals able to be processed in that time will be
	\beq
	\ntot=2\times10^{-4}\left(\frac{r_p^2}{92r_p+16\zeta}n^{2/3}\,\ell_\star\,\Delta\lambda\,D_t^2\,T\right)^{3/5}
	\eeq
	Here $\ell_\star=4\pi d^2\phi_\star$ is the stellar luminosity per frequency. However, if instead, the noise is dominated by the zodiacal contribution in eqn (\ref{tau}), it will scale as $d^4$, and\footnote{If both contributions are important, a septic equation must be solved that interpolates between these two behaviors along the lines of \cite{agol2007rounding}, but this will not be undertaken here.}
	\beq
	\ntot=6.6\times10^{-6}\left(\frac{n^{4/3}\, \ell_\star^2\, r_p^2\,\Delta\lambda\, D_t^4\,T}{\phi_z\,\lambda^2}\right)^{3/7}\label{scaling}
	\eeq
	This can be compared to the scaling $\ntot\sim 17.29(T/\text{yr})^{.41}-1.79$ found in \cite{stark2014maximizing}, which uses a sophisticated target selection algorithm and actual star catalogs: notice the exponent closely matches $3/7=.429$. They also find that the number scales with telescope diameter as $N\sim .39(D_t/\text{m})^{1.80}-.9$, which corresponds nicely with our value of $12/7=1.71$. 
	
	So, while the continuum approximation is not perfect for missions aiming at sample sizes of 10-100, it provides a worthwhile approximation to bear in mind. It suggests that to increase return, telescope area, bandwidth, and mission lifetime should be maximized, and noise minimized, but that the returns for all but telescope diameter will be sublinear. Additionally, as is already well known, it suggests to look in directions of higher stellar density if the survey is not to be full-sky, and to focus on intrinsically brighter planets.
	
	\subsection{Future Missions}
	
	\begin{table}
		\vskip.4cm
		\begin{center}
			\begin{tabular}{|c|c|c|c|}
				\hline 
				telescope & year & life/photo & tech \\
				\hline
				JWST & 2021 & 2  & - \\
				WFIRST & 2020s & 4 & - \\
				GMT & 2024 & 5 & - \\
				TMT & 2027 & 5 & - \\
				E-ELT & 2025 & 10 & - \\
				HabEx & 2030s & 12 & 12 \\
				OST (6 m)& 2035 & 10 & 10 \\
				OST (9 m)& 2035 & 20 & 20 \\
				LUVOIR (8 m) & 2038 & 56 & 56 \\
				LUVOIR (15 m) & 2038 & 108 & 108 \\
				RAVE/GAIA & 2018 & - & 8,365\\
				ELF (20 m) & - & 12 & - \\
				ELF (50 m) & - & 100 & 100 \\
				OWL-MOON & - & 1,000 & 1,000 \\
				Hypertelescope & - & 50,000 & 50,000 \\
				FOCAL & - & $4\times 10^6$ & $4\times10^6$ \\
				\hline
			\end{tabular}
		\end{center}
		\caption{The number of Earthlike planets around main sequence stars for which each transition could be measured with different proposed telescope technologies.}
		\label{Nearths}
	\end{table}
	
	Now, we comment on future exoplanet missions, estimate their yields, and comment on their various targets and constraints. The result of this is summarized in Table \ref{Nearths}.
	
	JWST will be a 6.5 meter space telescope with wavelength range .6-29 $\mum$ (\cite{jwst}). It has the coronagraphic sensitivity to detect Jupiter at 30 pc, but its rather high noise floor of 10 ppm will restrict its exoplanet targets to bright red dwarfs hosting large planets (\cite{greene2016characterizing}). Though it will have to be incredibly lucky to detect oxygen, JWST will be able to detect $\methane$ and $\carbon$ around Trappist-1 planets with 10 transits (\cite{krissansen2018detectability}) and around GJ876 (\cite{arney2017pale}) after 65 hours, so the expected yield is at least 2.  Its mid-IR capabilities make detecting larger molecules possible, such as are produced by technosignatures, though the sensitivity of these may only be feasible around white dwarfs (\cite{lin2014detecting}).
	
	WFIRST (\cite{wfirst}) is a future 2.4 meter infrared space telescope equipped with a coronagraph. Its noise floor will be a few ppb, and though it will detect thousands of exoplanets down to Mars mass, these will primarily be outside of the snow line of their system. In \cite{seager2018search} it was estimated that WFIRST will be able to detect atmospheric gases around 4 Earthlike exoplanets.
	
	ELTs are extremely large ground based telescopes slated for the 2020s. Their contrast is limited by Earth's atmosphere to be $10^{-8}$ in the near infrared, and so red dwarf stars will be their primary exoplanet targets. GMT (25 m) (\cite{gmt}) and TMT (30 m) (\cite{tmt}) are expected to yield 5-10 Earthlike planets, and can detect oxygen on an Earthlike planet orbiting an M4 star 5 pc away in 70 hours. The E-ELT (\cite{eelt}) will be 39 m and cover the .39-2.5 $\mum$ range at high spectral resolution. It will be able to target 10-20 rocky habitable zone planets that will have been flagged for follow-up by TESS, GMT and TMT (\cite{lopez2019detecting}).  In our analyses, we quote the pessimistic values of these numbers.
	
	
	
	HabEx (\cite{habex}) is a proposed 4 meter space telescope designed to directly image Earthlike planets around sunlike stars out to a distance of 8 pc. It includes a UV spectrograph, coronagraph, and is equipped with a starshade that can be used for the most interesting 50-100 systems. The integration time needed to detect the Earth-sun system at 7 pc is 1 month, and will spend 3.5 years detecting exo-Earth candidates, with an expected return of 12.  Because its frequency bands extend into the visible range, it will be capable of characterizing technosignatures from city lights at night.
	
	OST (\cite{ost}) is a proposed 9.1 m space telescope with active cooling down to 4 K that can resolve terrestrial planets from 5-660 $\mum$, primarily around M dwarfs. Though most of this range has a contrast of 1 ppm, the 25-200 $\mum$ range will be 1-2 orders of magnitude better than JWST. Over its 5-10 year lifetime goal, it will be able to detect ice features, NH${}_3$, and, thanks to its coronagraph with $10^{-7}$ contrast sensitivity in the infrared, potentially technosignatures. It is projected to measure CO${}_2$ and O${}_3$ on 30 and 20 rocky planets, respectively. An alternative design consists of a 5.9 m primary mirror with the same science goals but correspondingly decreased planet yield, which we infer using eqn (\ref{scaling}) (\cite{ost2020}).
	
	LUVOIR (\cite{luvoir}) is a proposed mission that will be capable of directly imaging the Earth, Venus and Jupiter at a distance of 13 pc. The diameter will be either 8 or 15 meters, depending on budget choices, and will carry multiple instruments capable of observing in the 0.1-2.5 $\mum$ range. It would have a prime mission lifetime of 5 years, with a lifetime goal of 25 years. The 15 meter design has an expected return of 54 Earthlike planets around AFGKM stars after its initial survey, and the 8 meter design anticipates 28. Using our scaling from eqn (\ref{scaling}), this translates into an expected 108 and 56 Earthlike planets over the full 25 year lifespan, respectively.
	
	The RAVE DR5 and GAIA DR1 data sets were searched for civilizations using a significant fraction of stellar light in \cite{zackrisson2018seti}, which could in principle alter the star's spectrum and luminosity profile. There, 8,365 stars in both catalogs were selected by comparing parallax distance with IR spectra. Though 6 potentially anomalous stars were found, these can all be explained by measurement error and, in one case, a binary companion.
	
	ELF (\cite{elf}) is a potential ground-based circular array of 9-25 4-8 meter telescopes surveying in the 0.3-5 $\mum$ range, and with polarization capabilities. It will be able to directly image nearby planets with continent-scale resolution, and the larger version would be able to detect waste heat from a civilization that uses 25 times our current energy output. It will be able to detect CFCs in about a day's exposure time.  The proposal estimates that at least a dozen Earthlike planets will be characterizable with a 20 m dsign, and over 100 with a 50 m design.
	
	OWL-MOON (\cite{owlmoon}) is a potential design that would place an overwhelmingly large telescope in a crater on the south pole of the moon. It could potentially be 50-100 m large (or even larger), and would sidestep conventional ground based telescope challenges such as atmospheric noise, thermal noise, and wind stresses. It would be able to resolve emission lines from an Earth-Sun system at 40 pc in 3 hours, and the total estimated yield for the size given above is quoted as 1,000.
	
	The hypertelescope (\cite{hypertelescope}) is a potential `flotilla' array of interferometers potentially spanning 100 km across. This gives it an effective collecting area comparable to a 39 meter telescope. With this setup, a 30 minute exposure would be able to resolve the spectrum of an Earthlike planet at 3 pc for a 30$\times$30 pixel grid. This would enable detection of continents and cities at night, should any be present in this distance range. Extrapolating the yields of OWL-MOON to a 1 km design using eqn (\ref{scaling}), we find that even a modest hypertelescope would yield 50,000 Earthlike planet spectra.
	
	FOCAL (\cite{focal}) is a potential plan to put a telescope to 550 AU and beyond in order to use the sun's gravitational lens to vastly enhance the collecting area. With this, a 1 meter telescope would be able to reposition itself to scan over the image of a planet, giving it an effective diameter of 12 Earth radii. Over a 7 week integration time for an Earthlike planet at 30 pc, it would be able to create a megapixel image, which, to reiterate, corresponds to 1000$\times$1000 pixels. Observing multiple targets with this technique would be challenging, as repositioning would be prohibitive. To estimate the total number of exo-Earths that could be detected with this technology, we tally the total number within a 300 pc radius of the sun.
	
	Having overviewed the telescopes slated for deployment in the near (and not so near) future, we now outline the general formalism for how well each quantity can be measured for a given survey size. We make our analysis as analytic as possible, in order to track the dependence of the uncertainties on the various experimental parameters as clearly as possible. We illustrate our formalism with several idealized test cases first, in order of increasing complexity: in section \ref{sing}, we begin with analyzing the case where only one biosignature is measured, and find that maximizing science return in this case is indeed equivalent to maximizing the total number of observed systems, as is so often stated in the literature. In section \ref{mult} we extend our analysis to the measurement of multiple biosignatures, and explore how to combine two different datasets in our framework. Depending on the precise setup, we find recommendations for optimum survey size. In section \ref{conf} we incorporate false positives and false negatives, and study how these affect our formalism. We find that the desired degree of confidence will generically be of order the observed fraction of systems that possess a given biosignature, in a variety of particular setups.
	
	\section{Single Biosignature}\label{sing}
	
	To begin, we make several simplifications: first, we treat the stellar population as identical, having no characteristics that would affect its probability of hosting a biosphere, or our ability to detect it. Secondly, we assume all biosphere detections are unambiguous. If we have the ability to survey $\ntot$ planets and the fraction of planets with detectable biospheres is $\fbio$, then the number of detections $\ndet$ is given by a binomial distribution:
	\beq
	p(\text{data}|\fbio)={\ntot\choose \ndet}\,\fbio^{\ndet}\left(1-\fbio\right)^{\ntot-\ndet}\label{pB}
	\eeq
	When needed, this distribution will be referred to as $B(\ntot,\ndet,f)$ below, and $\text{data}$ refers generically to the numbers of total and detected systems relevant to the consideration at hand- here, $\ndet$ and $\ntot$.
	
	We will be more interested in determining the ratio $\fbio$, given the number of surveyed planets and detected biospheres. This is related to the above through Bayes' Theorem,
	\beq
	p(\fbio|\text{data})\,\propto\, p(\text{data}|\fbio)\,p_\text{prior}(\fbio)
	\eeq
	As an example, let us take the case where $p_\text{prior}(\fbio)$ is given by a uniform distribution. Then the probability distribution for $\fbio$ will be described by a beta distribution,
	\beq
	p(\fbio|\text{data})=\beta(\ntot,\ndet,\fbio)\label{pbeta}
	\eeq
	where we define $\beta(\ntot,\ndet,\fbio)=(\ntot+1)B(\ntot,\ndet,\fbio)$
	The expected value and variance of $\fbio$ are given by (see, for example, \cite{johnson1995chapter}):
	\beq
	\langle \fbio\rangle=\frac{\ndet+1}{\ntot+2},\quad \sigma_{\fbio}^2=\frac{\left(\ndet+1\right)\left(\ntot-\ndet+1\right)}{\left(\ntot+3\right)\left(\ntot+2\right)^2}\label{avg}
	\eeq
	In the limit where both $\ndet$ and $\ntot$ are large we have, defining the observed fraction as $\rdet=\ndet/\ntot$,
	\beq
	\fbio\sim \rdet\pm \sqrt{\frac{\rdet\left(1-\rdet\right)}{\ntot}}
	\eeq
	As expected from generic properties of the beta distribution. This informs us that to get a precise measure of the probability of hosting a biosphere, the total number of surveyed planets should be larger than $1/\rdet$.
	
	If we fail to detect any biospheres after our survey, then the probability distribution for $\fbio$ reduces to
	\beq
	p(\fbio|0)=\left(\ntot+1\right)\left(1-\fbio\right)^{\ntot+1}
	\eeq
	This tends to 0 for $\fbio\gtrsim1/\ntot$. Then, we could infer that $\fbio\sim(1\pm1)/\ntot$.
	
	If we want to ensure that we have probability $p_0$ of detecting at least one biosphere, we will need to design an experiment capable of surveying at least
	\beq
	\ntot>\frac{\log\left(1-p_0\right)}{\log\left(1-\fbio\right)}
	\eeq
	different planets. In the limit $\fbio\ll1$, this reduces to $\ntot\gtrsim\mathcal O(1)/\fbio$, where the coefficient depends on the desired confidence.
	
	\subsection{Priors}
	We now address the question of what effect the prior distribution $p_\text{prior}(\fbio)$ has on the final inference. As usual, if enough samples are taken, the form of the prior is diminished; however, the expected return will likely not be quite large enough to completely extinguish the prior's influence. Here, we investigate several alternatives to a uniform prior, and the effects they introduce.
	
	As a first example, let us suppose that the prior distribution for $\fbio$ is log-uniform, $p_\text{prior}(\fbio)\propto1/\fbio$. For the most part, this does not substantially alter the analysis: the result is simply a shift in the parameters of the beta distribution, so that $p(\fbio|\text{data})=\beta(\ntot-1,\ndet-1,\fbio)$. The expected value of $\fbio$ and variance in this case are given by
	\beq
	\langle\fbio\rangle=\frac{\ndet}{\ntot+1},\quad \sigma_{\fbio}^2=\frac{\ndet\left(\ntot-\ndet+1\right)}{\left(\ntot+2\right)\left(\ntot+1\right)^2}
	\eeq
	These numbers can be seen to be shifts of those in eqn (\ref{avg}). These results hold as long as at least one detection is made. If $\ndet=0$, however, it is necessary to introduce the smallest conceivable probability of a planet hosting a biosphere $\pmin$ in order to regulate the otherwise divergent expressions. This gives
	\beq
	\fbio\sim \frac{-1\pm1}{\ntot\left(\gamma_E+\log\left(\pmin\,\ntot\right)\right)}
	\eeq
	where $\gamma_E$ is the Euler-Mascheroni constant, and we have taken the large $\ntot$, small $\fbio$ limit prior to displaying these formulas. For this to be positive, $\pmin<.56/\ntot$, which is a sensible condition anyway if no life had been detected by that point. Similarly, using the Jeffreys prior, $p_\text{prior}(\fbio)\propto1/\sqrt{\fbio(1-\fbio)}$, yields $p(\fbio|\text{data})=\beta(\ntot-1,\ndet-1/2,\fbio)$, again tantamount to simply shifting the survey size and number of detected signals by small numbers.

\begin{centering}
		\begin{figure*}
			\centering
			\includegraphics[width=.6\textwidth]{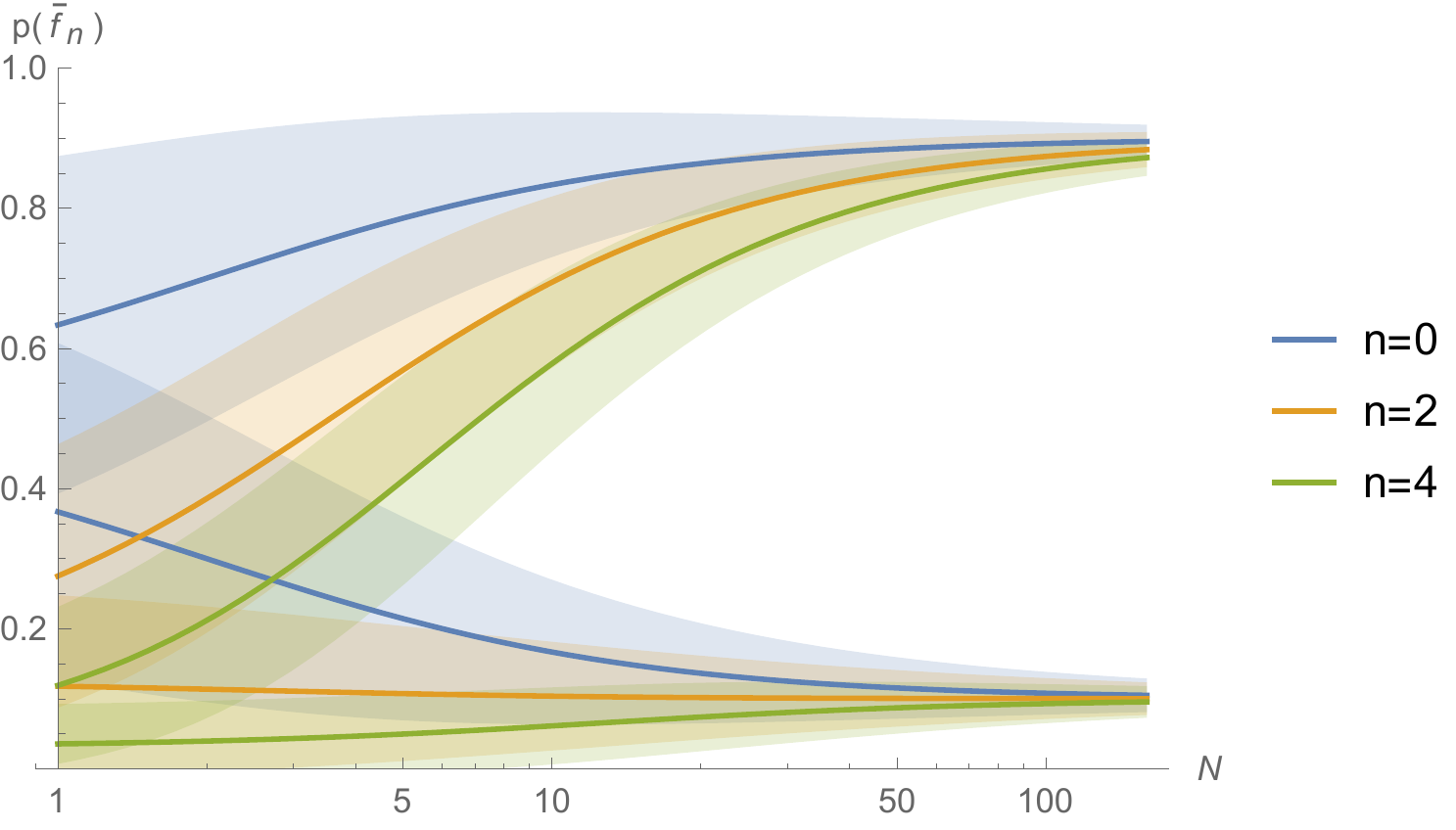}
			\caption{Inferred value of $\bar f_n$ (with one standard deviation width) for various values of number of compounded steps $n$, as given by eqn (\ref{nstep}). Two curves are displayed for each: one with the fraction of systems that display the signal equal to .1, the other with .9.}
			\label{logs}
		\end{figure*}
	\end{centering}

	Let us also note an important factor when measuring more advanced biosignatures: because what we often measure will be products of sequential probabilities, each of which we take to be uniform, the prior we should assign to the measured quantity will not in fact be uniform. If we measure $\bar f_n=\prod_{k=1}^nf_k$, then the prior distribution for this can be found by successively integrating over the latent variables, assuming a uniform distribution for each:
	\beq
	p_\text{prior}(\bar f_n)=\frac{1}{(n-1)!}\log\left(\bar f_n^{-1}\right)^{n-1}
	\eeq
	So that, not only will the inferred distribution depend on the prior, but also on the number of steps we choose to include in our counting scheme. As the number of steps to include is far from clear, this makes the desire to collect enough samples to circumvent this issue quite strong. But how many is enough?  For this, we compute the mean and variance by integrating eqn (\ref{pB}) with the above prior, and take the large $\ntot$ limit:
	\beq
	\langle\bar f_n\rangle \rightarrow r+\frac{1-2 r}{\ntot}+\frac{(n-1)(1-r)}{\log(r)\ntot}\label{nstep}
	\eeq
	Where $r=\ndet/\ntot$. The first term in the $\mathcal{O}(1/\ntot)$ correction is present even for a uniform prior, and simply results from an expansion of eqn (\ref{avg}). The term proportional to $n$ represents a systematic negative shift of the average value of $\bar f_n$, arising from the preference for smaller compound probabilities. The variance, however, is unchanged to leading order in $\ntot$: $\sigma^2\rightarrow r(1-r)/\ntot$. The number of samples needed for the negative shift to be within a standard error is then
	\beq
	\ntot>n^2\frac{1-r}{r\log(r)^2}
	\eeq
	This diverges for $r$ close to 0 or 1, and attains a minimum of $1.54n^2$ at $r=.20$. The dependence of the mean and standard deviation on $n$ and $\ndet$ are displayed in Fig. \ref{logs}.
	
	For the remainder of this paper, we will only report on the leading behavior of these quantities.

	\section{Multiple Nested Biosignatures}\label{mult}
	
	Here we extend the results of the previous section, which dealt with the detection efficiency of a single biosignature, to the case where a number of different biosignatures are observed. Our chief concern will be to determine how well we will be able to measure the states of various exoplanet biospheres: for the various transitions discussed in section \ref{sign}, we wish to determine the transition rates for each level, the extent to which they can be measured with a given technology, and a means to inform policy recommendations for instrument design in order to maximize the scientific profit toward this goal. If the returns for the various biosignatures are forecast for a specific mission, as is collated for several upcoming telescopes in Table \ref{Nearths}, the final numbers may be easily imported into the expressions we derive to find the achievable measurement accuracies.
	
	The quantities we wish to determine with as much precision as possible are $f_i$, the fraction of planets that, having achieved level $i-1$, also achieve level $i$. As before, the setup will be to assume that we have the ability to survey a number $\ntot$ of systems. Then, the total number of observed planets of each types will be given by a conditional binomial distribution:
	\beq
	p(\text{data}|\{f_i\})=\prod_i B(N_{i-1},N_i,f_i)
	\eeq
	where $N_0=\ntot$, $\{f_i\}$ refers to the full collection of inferred fractions, and we have made the simplification that the detection efficiency of each biosignature is perfect (to be discussed further in section \ref{conf}). The crucial feature of this distribution is that aside from the nested dependence on the number of available systems at each level, the distribution of each variable functions independently of the others. Because of this, inverting this to yield a likelihood function for the variables $f_i$ given the observational yield $N_i$, we have simply
	\beq
	p(\{f_i\}|\text{data})=\prod_i\beta(N_{i-1},N_i,f_i)\label{multibeta}
	\eeq
	The normalization assumes that the $f_i$ are distributed uniformly, and eqn (\ref{avg}) may be used to compute
	\beq
	\langle f_i\rangle=\frac{N_i}{N_{i-1}},\quad\sigma_{f_i}^2=\frac{\frac{N_i}{N_{i-1}}\left(1-\frac{N_i}{N_{i-1}}\right)}{N_{i-1}}
	\eeq
	
\begin{centering}
		\begin{figure*}
			\centering
			\includegraphics[width=.6\textwidth]{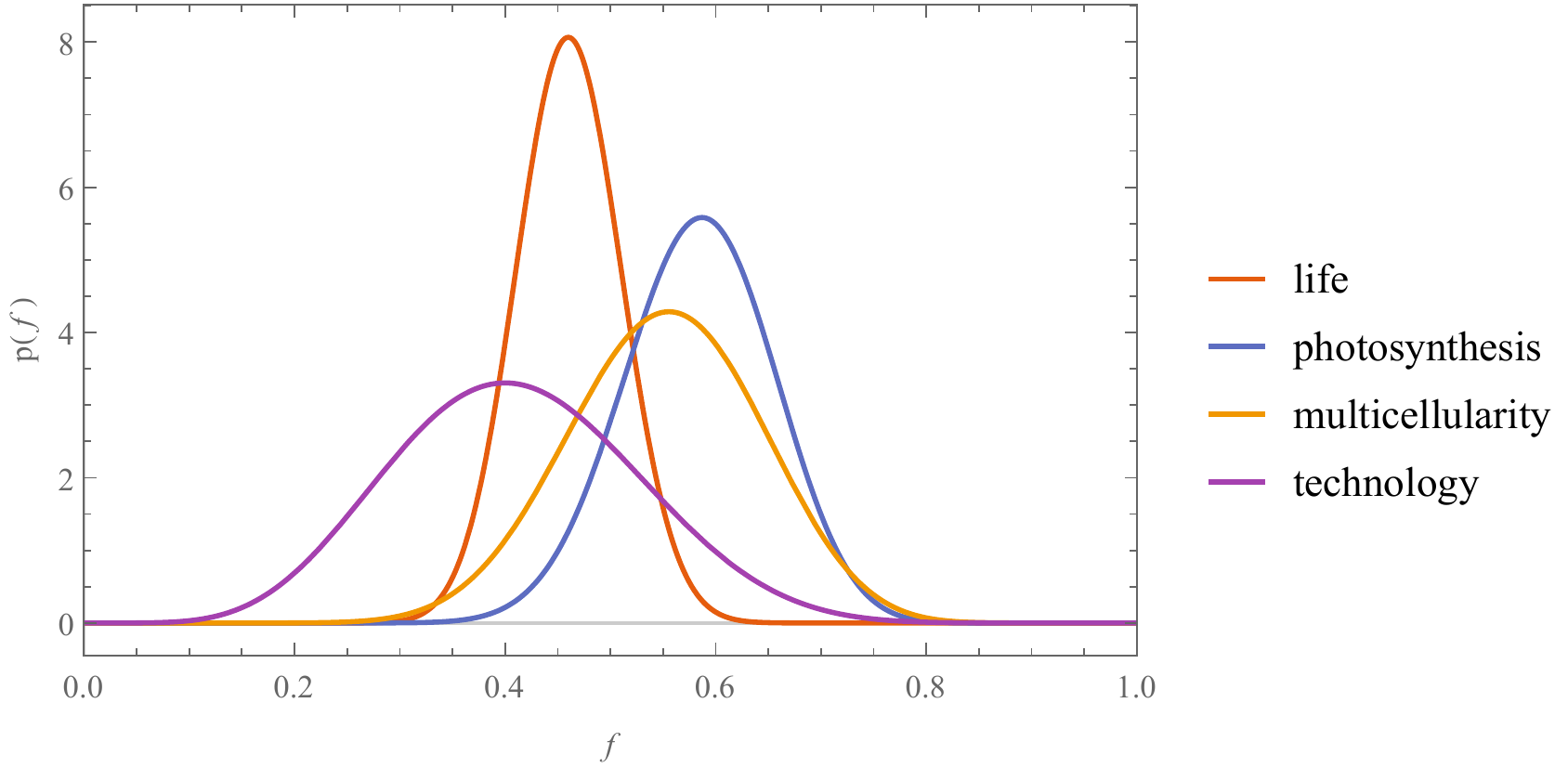}
			\caption{Inferable distributions for $ f_i$s under a random realization of 100 total samples. Here, the distributions are given by eqn (\ref{multibeta}), with expected values equal to 1/2 for each, and the number of systems possessing each successive quality are 46, 27, 15, and 6. Note that the uncertainties grow for each successive biosignature fraction. It is worth stressing that in this figure, it is assumed that the detection of the biosignature would unambiguously signal the presence of the stage of life under consideration. This assumption will be relaxed in section \ref{conf}.}
			\label{f100}
		\end{figure*}
	\end{centering}

	The most important feature of these expressions is that the variance of each quantity is controlled by the number of systems displaying the previous signal. A consequence is that, as long as the probability of attaining each successive transition is not 1, the confidence in measuring each successive variable will diminish. This is illustrated in Fig. \ref{f100} for the case where all $f_i=1/2$, and for a total sample of 100 stars, as would be appropriate for LUVOIR or ELF, from Table \ref{Nearths}. Here, while the measurement of $f_\text{life}$ can be determined with reasonable accuracy, the latest stages of innovation are highly dominated by sample noise.
	
\begin{centering}
		\begin{figure*}
			\centering
			\includegraphics[width=.6\textwidth]{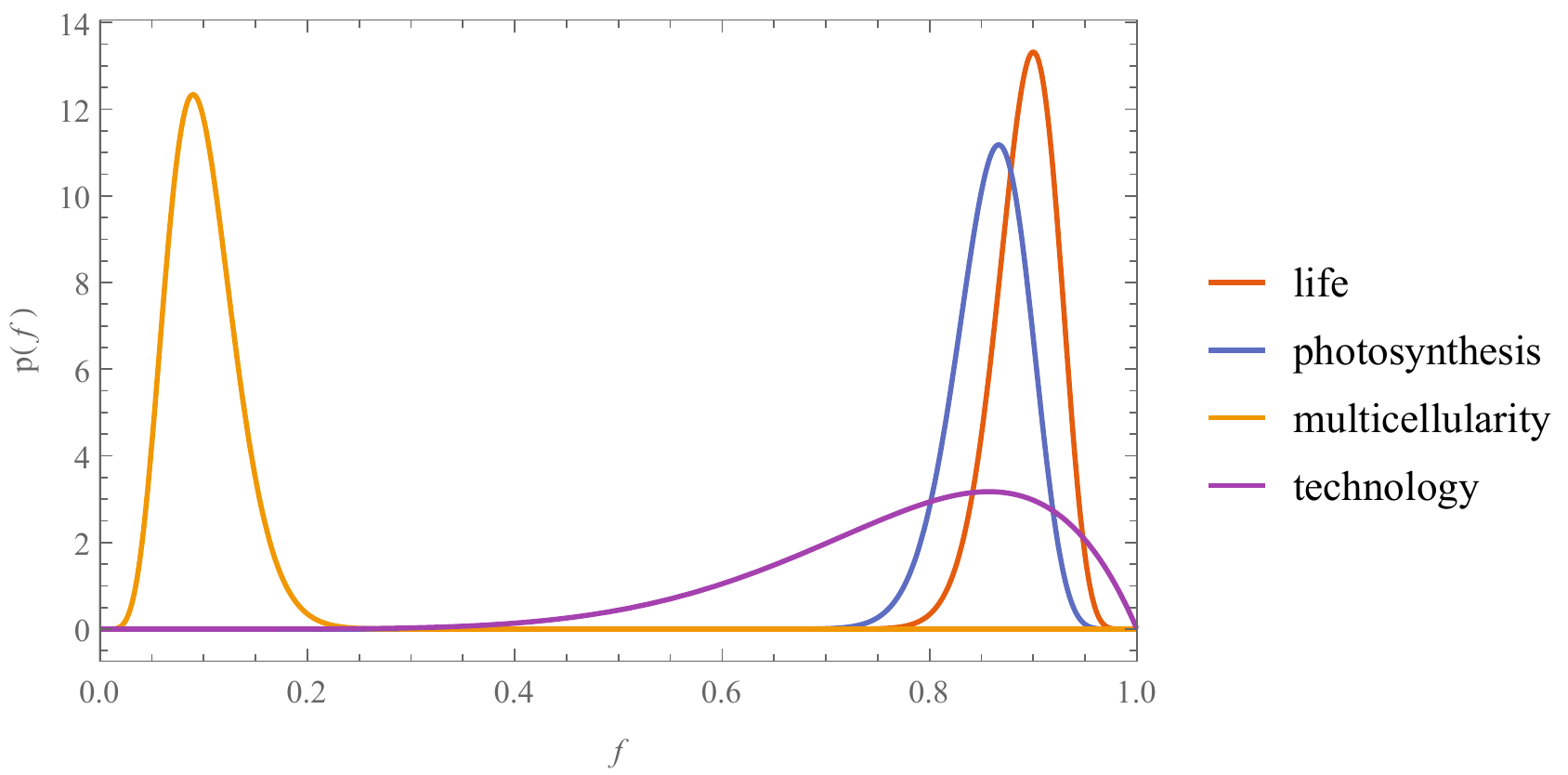}
			\caption{The same as above, except taking all $ f_i=.9$ except for $ f_\text{mult}=.1$. The number of systems with each quality in this sample are 90, 78, 7, and 6. Here, the uncertainties for the first three biosignatures are roughly the same, but the uncertainty for the last is very large, as it is dictated by the small number that precedes it.}
			\label{f91}
		\end{figure*}
	\end{centering}

	This may be contrasted to another possible case displayed in Fig. \ref{f91}, where all but one transition is taken to occur rather frequently, $ f_i=.9$, leading to one bottleneck in the road of progress which we have chosen for purely illustrative purposes to be $ f_\text{mult}=.1$. From here, it can be seen that for all transitions up to and including the bottleneck, the underlying fraction of systems can be measured quite accurately. The transitions which occur after the bottleneck, however, are rather poorly constrained.
	
	However, as the initial sample size is increased, the remainder after the bottleneck transition can attain statistically significant values. This suggests that if we do expect one of the transitions to be a bottleneck, we should not invest much effort in designing a mission to measure the transition probabilities that occur after this until we are able to pool from a large enough number of systems to beat the sample noise. From above, the number required will be
	\beq
	\ntot\geq\frac{1}{\prod_{i=1}^{b} f_i}
	\eeq
	Then we will not do well to invest in measuring $f_b$ until enough systems have been harvested so that this equality is satisfied. For any value of total sample given in Table \ref{Nearths}, this defines values for the $f_i$ for which this measurement will be worthwhile.
	
	Alternatively, one strategy will be to design a supplemental mission capable of measuring the effects of some transition, but not the previous targets, provided that the survey is much larger than the one originally under discussion. Such a mission would only be able to measure the compound probability $\prod_{i=1}^b f_i$, but supplementing with information gleaned from the original mission on each individual transition rate before that will enable us to disentangle the separate effects.
	
	As it stands now, though, we do not know the individual $f_i$s, which were needed in our heuristic for the threshold number of systems to observe. It would serve us well, then, to determine the most primitive of these values first, and then work our way through the succession as more information becomes available. Only through this method will we be able to accurately determine the expected yields of future missions with any sort of certainty.
	
\begin{centering}
		\begin{figure*}
			\centering
			\includegraphics[width=\textwidth]{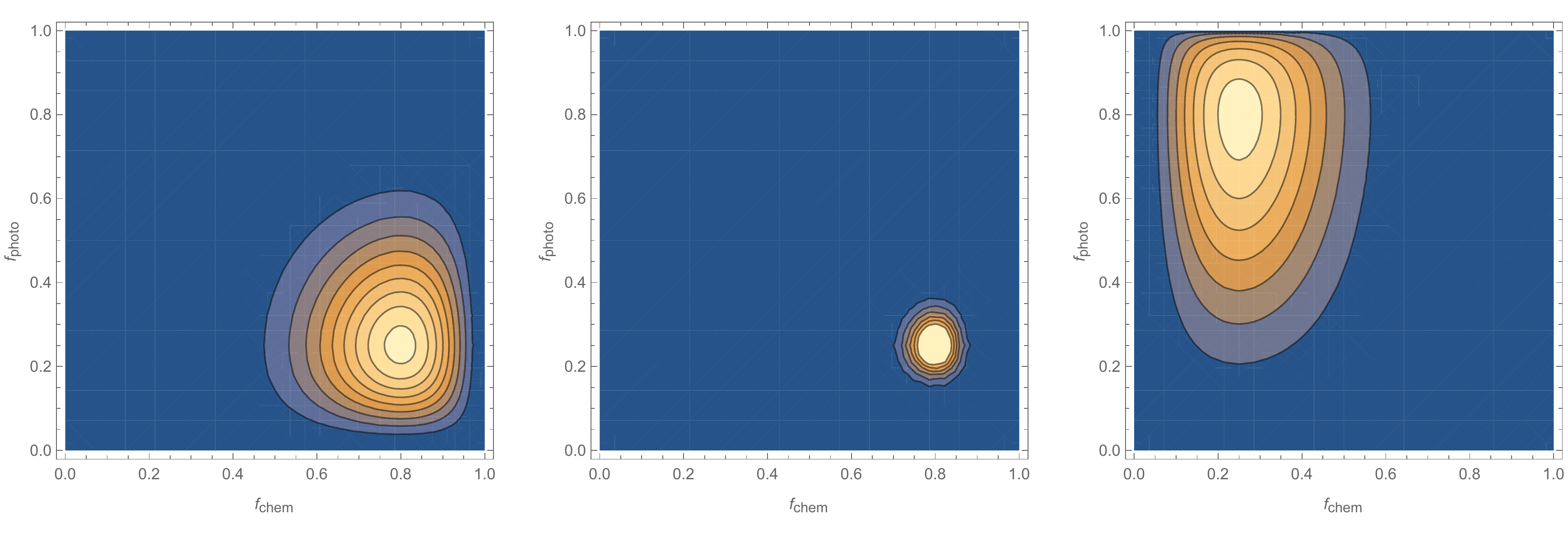}
			\caption{Three different likelihood functions, superposed, illustrating the range of uncertainties on these parameters that will be possible with the upcoming surveys. The distributions are given by eqn (\ref{multibeta}), with values taken to be $(f_\text{life}=.8,f_\text{photo}=.25,\ntot=10)$, $(f_\text{life}=.8,f_\text{photo}=.25,\ntot=100)$, and $(f_\text{life}=.25,f_\text{photo}=.8,\ntot=10)$. Though the fractions are simply swapped between the first and third plots, the uncertainties are magnified in the third, reflecting the diminution of successive samples in the latter case.}
			\label{both100}
		\end{figure*}
	\end{centering}

	We also display a two-dimensional joint pdf in Fig. \ref{both100}, so as to give a feel for the correlations between variables.
	
	\subsection{Combining Two Data Sets}\label{comb}
	
	We wish to discuss the scenario where incomplete data on each of the desired signals is gained, and the limits on what we can infer for each of the $f_i$s. Our treatment will be far from comprehensive, but we will illustrate several examples of situations where we are likely to find ourselves in the future. For simplicity, we restrict ourselves here to just two signals to be measured, $f_1$ and $f_2$, where transition 1 precedes number 2 in our ordering. For example, we could be measuring the fraction of planets which attain life and photosynthesis.
	
\begin{centering}
		\begin{figure*}
			\centering
			\includegraphics[width=\textwidth]{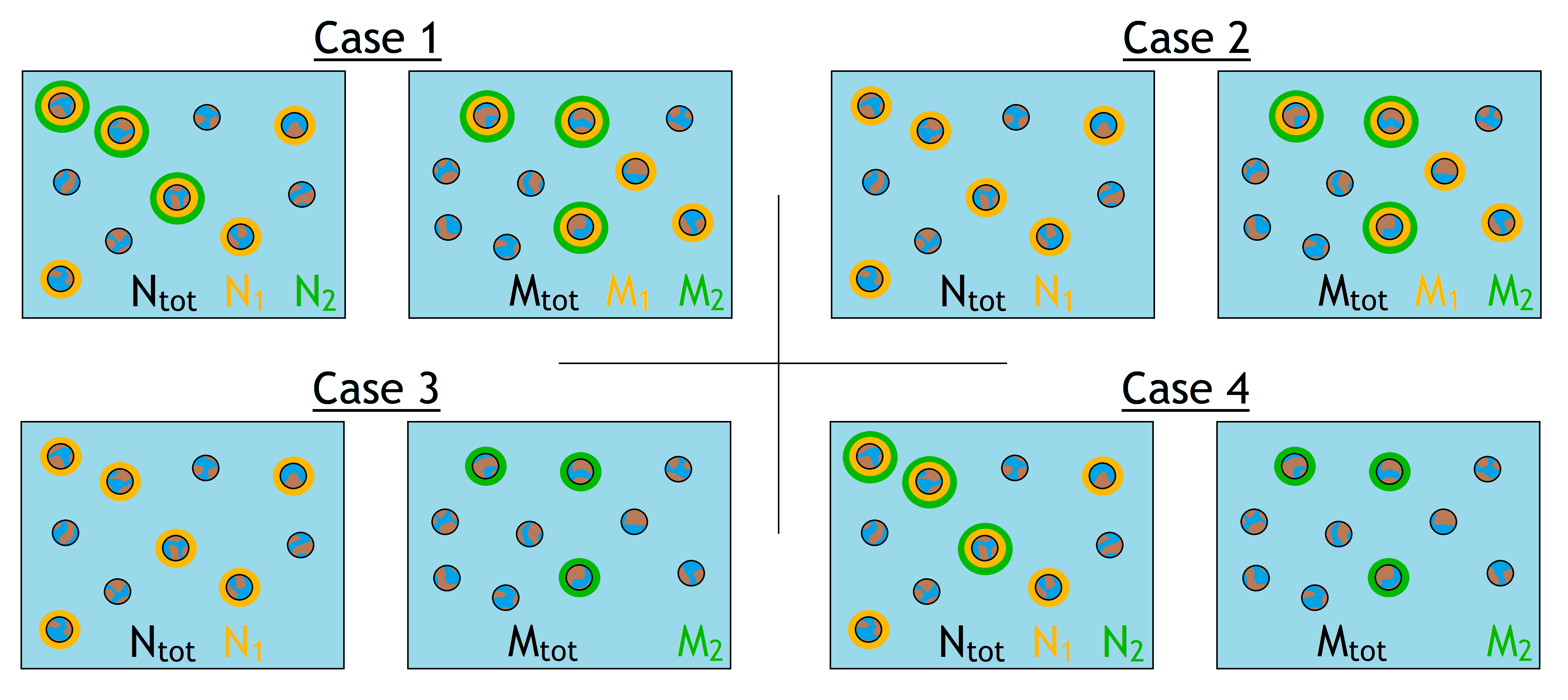}
			\caption{Different possible cases when combining two data sets. Here, the orange glow indicates the presence of a more primitive signal, and the green a more advanced.}
			\label{cases}
		\end{figure*}
	\end{centering}

	Suppose we take a population $\ntot$ with $N_1$ instances of signal 1 and $N_2$ instances of signal 2, and similarly a separate population $\mtot$ with measurements $M_1$ and $M_2$. This may come about from data from two different telescopes, for instance, or else from data from the same telescope where we expect not to be able to measure some of the signals in each population. Let us go through several different possible cases, which are illustrated in Fig. \ref{cases}.
	
	{\bf Case 1: All present}
	
	This case is quite easy to handle: one can simply combine the two datasets into one larger one. The inferred distributions of the parameters is then
	\begin{eqnarray}
	p(f_1|\text{data})=\beta(\mtot+\ntot,M_1+N_1,f_1),\nonumber\\
	p(f_2|\text{data})=\beta(M_1+N_1,M_2+N_2,f_2)
	\end{eqnarray}
	the analysis of the previous section can be straightforwardly applied to derive the mean and variance of these two quantities. The means are $\langle f_1\rangle=\hat r_1$ and $\langle f_2\rangle =\hat r_2/\hat r_1$, and, more importantly for our analysis, the variances are.
	\beq
	\sigma_1^2=\frac{\hat r_1\left(1-\hat r_1\right)}{\mtot+\ntot},\quad\sigma_2^2=\frac{\frac{\hat r_2}{\hat r_1^2}\left(1-\frac{\hat r_2}{\hat r_1}\right)}{\mtot+\ntot}
	\eeq
	Where $\hat r_i=(M_i+N_i)/(\mtot+\ntot)$. As can be seen, $\mtot$ and $\ntot$ here are on an equal footing, so that in order to maximize the desired signal, one could choose to increase either quantity, giving preference to whichever would be cheapest.
	
	{\bf Case 2: $N_2=0$}
	
	In this case, we have information about the more primitive signal in both data sets, and the more advanced signal only in one. This scenario could arise, for instance, when combining data from two different telescopes, such as JWST and WFIRST, as displayed in Table \ref{Nearths}. Alternatively, this situation could be relevant if measurement of the second signal is impossible for part of the sample from a single telescope, due to faintness or other factors. For this, the probability distributions are rather simple to write down:
	\beqa
	p(f_1|\text{data})=\beta(\mtot+\ntot,M_1+N_1,f_1),\nonumber\\ p(f_2|\text{data})=\beta(M_1,M_2,f_2)
	\eeqa
	The statistics for $f_1$ are the same as in case 1, but now $\langle f_2\rangle=r_2/r_1$, and
	\beq
	\sigma_2^2=\frac{r_2\left(1-\frac{r_2}{r_1}\right)}{r_1^2\,\mtot}
	\eeq
	This is usually going to be larger than the variance of $f_1$ unless $\langle f_2\rangle$ is either very close to 0 or 1. As such, in this case priority should be given to increasing $\mtot$ rather than $\ntot$: this will, after all, improve measurements of both quantities, rather than just one.
	
	{\bf Case 3: $N_2=0$, $M_1=0$}
	
	In this case, we have information about the first transition from the first data set, but the second data set only gives us information about the product $f_1\, f_2$. This scenario could come about with the combination of any of the first five entries delineated in Table \ref{Nearths} with the technosignature search results of RAVE/GAIA. In this case, we have
	\beqa
	p_1(f_1|\text{data})=\beta(\ntot,N_1,f_1),\nonumber\\ p_{12}(f_1f_2|\text{data})=\beta(\mtot,M_2,f_1f_2)
	\eeqa
	In order to reconstruct the desired distribution $p_2(f_2|\text{data})$, we integrate over $f_1$:
	\beq
	p_2(f_2|\text{data})=\int_0^1df_1\, p_1(f_1|\text{data})\,p_{12}(f_1 f_2|\text{data})\label{case3}
	\eeq
	
	This integral, along with most of the subsequent ones we will encounter, can be expressed exactly in terms of hypergeometric functions, essentially by definition. However, these are rarely illuminating and often difficult to manipulate, so we choose not to display them explicitly unless they are unavoidable for our conclusions. Note that we do not include the extra Jacobian factor of $f_2$, and instead utilize the standard measure $df_1 df_2$. This is the only appropriate prescription, as the other would break the degeneracy between $f_1$ and $f_2$ in the $\ntot=0$ case.
	
\begin{centering}
		\begin{figure*}
			\centering
			\includegraphics[width=.6\textwidth]{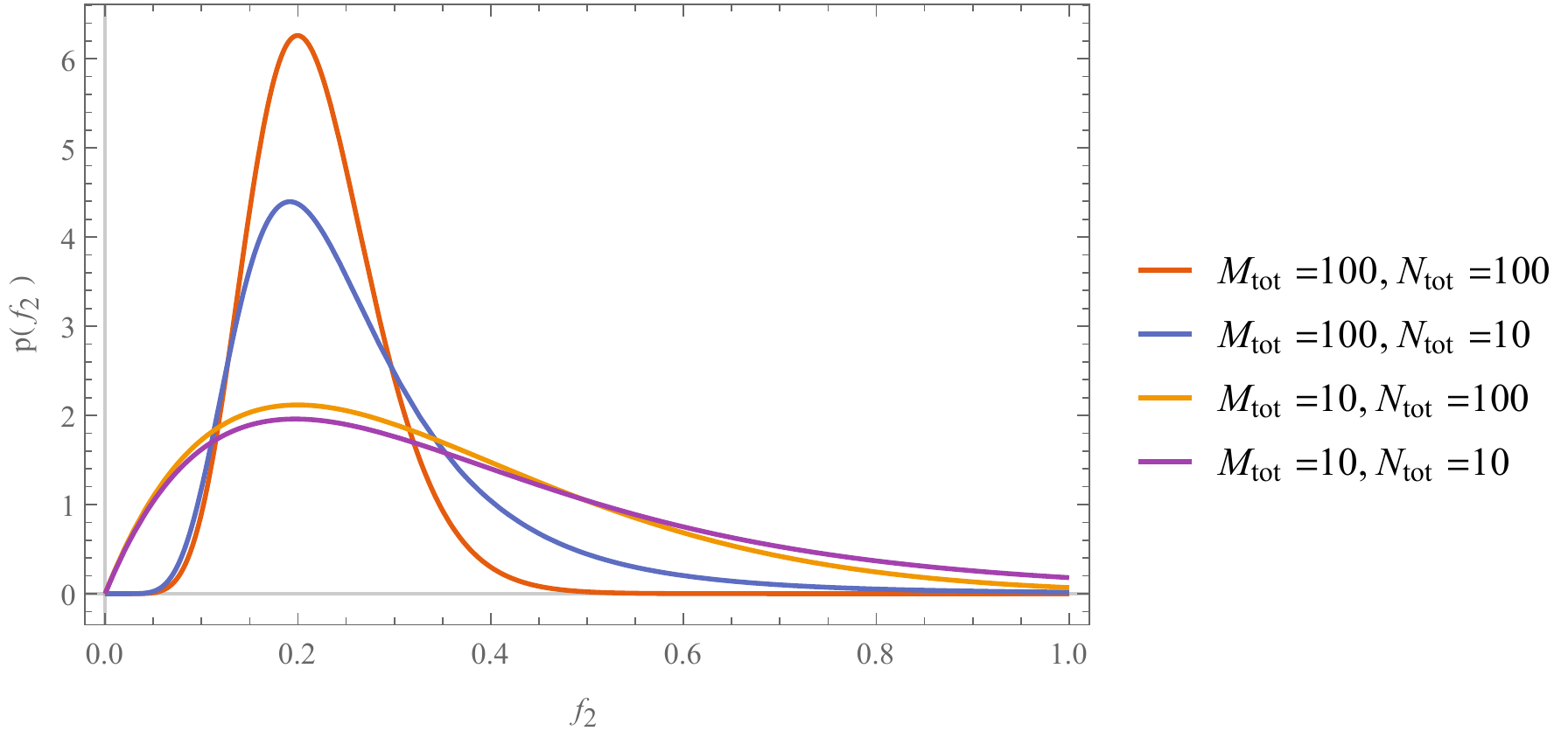}
			\caption{The distribution for $f_2$ in case 3, as given by eqn (\ref{case3}). Here, we have set $f_1=1/2$ and $f_2=1/5$. Notice that the variance is rather insensitive to $\ntot$, but is rather sensitive to $\mtot$.}
			\label{given}
		\end{figure*}
	\end{centering}

	In order to estimate the total variance for this distribution, it is useful to note that in the large sample limit, beta distributions can be well approximated by Gaussians:
	\beq
	\beta(\ntot,N_1,f)\rightarrow \mathcal N(r_1,\sigma_1)
	\eeq 
	where $r_1=N_1/\ntot$ and $\sigma_1^2=r_1(1-r_1)/\ntot$. Then eqn (\ref{case3}) becomes (assuming $p_1(f_1)$ is well localized away from 0 or 1 and that $r_{12}=M_2/\mtot<r_1$ for compatibility):
	\beq
	p(f_2|\text{data})\rightarrow \frac{e\char`\^\left(-\frac{(r_1 f_2-r_{12})^2}{2(f_2^2\sigma_1^2+\sigma_{12}^2)}\right)}{\sqrt{2\pi\,(f_2^2\sigma_1^2+\sigma_{12}^2)}}
	\eeq
	This is plotted for several different survey totals in Fig. \ref{given}. The variance of this distribution is approximately
	\beq
	\sigma_2^2\rightarrow \frac{r_{12}^2(1-r_1)}{r_1^3\,\ntot}+\frac{r_{12}(1-r_{12})}{r_1^2\,\mtot}\label{nmsigma}
	\eeq
	If $r_{12}=r_1$, corresponding to $f_2\approx1$, then the two contributions to the variance will be equal when the two surveys are the same size. Otherwise, the second term will dominate. Correspondingly, one should prefer to make $\mtot$ much larger than $\ntot$. The two contributions are equal when
	\beq
	\frac{\ntot}{\mtot}=\frac{r_{12}(1-r_1)}{r_1(1-r_{12})}
	\eeq
	An ideal mission would be designed so that the survey sizes match this ratio. This advice is only useful if there is some prior information about $r_1$ and $r_{12}$ beforehand; otherwise, the mission needs to be completed before it can be optimized. In the absence of this information, one may as well use the expectation value of this quantity: if $r_1$ and $r_2=r_{12}/r_1$ are both uniform, then the mission should be designed so that
	\beq
	\frac{\ntot}{\mtot}=2-\frac{\pi^2}{6}=.36
	\eeq
	
	Let us also note that, specifically when combining exoatmosphere searches with technosignature surveys, the number yields of the latter will vastly outnumber the former, at least for the foreseeable future.  In this case, $\mtot\gg\ntot$, and the first term in eqn (\ref{nmsigma}) will dominate as long as $f_\text{tech}>\ntot/\mtot$.  Taking the projection for the E-ELT yields, this threshold is .001.
	
	The above assumed that the probability distributions are well approximated by a Gaussian, which holds true if the observed values are well separated from 0.  In the current situation, with the absence of technosignature detections, this approximation does not hold, and instead we have $p_{12}(f_1f_2)\sim \mathcal{U}(0,1/M_2)$, where $\mathcal{U}(0,t)$ is a uniform distribution.  When the integral in eqn (\ref{case3}) is performed it results in $p_2(f_2)\sim M_2c_1(\text{min}(1,1/(f_2M_2)))\sim \mathcal{U}(0,1/M_2)$, irrespective of the distribution of $f_1$.  In this case, we recover no additional information about the fraction of life that develops into technological societies by measuring the prevalence of life.
	
	{\bf Case 4: $M_1=0$}
	
	The most difficult case to handle is when the more primitive signal is unable to be measured in one dataset. Again, this situation may arise from one telescope if the signal is unavailable for part of the sample, or else by the combination of data from two different telescopes. In this case, the data do not allow for the distributions of the two fractions to be factorized, so the full joint distribution of $f_1$ and $f_2$ must be employed:
	\beq
	p(f_1,f_2|\text{data})\propto B(\ntot,N_1,f_1)\,B(N_1,N_2,f_2)\,B(\mtot,M_2,f_1\,f_2)
	\eeq
	As in case 3, the variance can be estimated in the large survey size limit, when all binomial distributions can be well approximated by Gaussians. For the variance we find
	\beq
	\sigma_2^2\rightarrow\left(\frac{1}{\frac{r_{12}^2(1-r_1)}{r_1^3\,\ntot}+\frac{r_{12}(1-r_{12})}{r_1^2\,\mtot}}+\frac{1}{\frac{r_2(1-r_2)}{r_1\,\ntot}}\right)^{-1}
	\eeq
	And a similar expression for $\sigma_1^2$, with labels interchanged $1\leftrightarrow2$. This combines the variance from case 3 with a new term ``in parallel'', and so the total variance will resemble the smallest of the terms. Usually, this will be most readily given by the new term, and so the recommendation for this scenario is to increase $\ntot$ as much as possible. Again, the intuitive reasoning behind this is that increasing this gives information on both signals, whereas increasing $\mtot$ only gives information about $f_2$.
	
	The rest of the possible cases omit even more measurements from the survey and so are trivial. Having gained some general insights into how to maximize return when combining two datasets, we now turn our attention to the effects of imperfect surveys.
	
	\section{False Positives, False Negatives, and False Samples}\label{conf}
	
	Up to this point, we have operated under the simplification that any biosignature detection will be able to be unambiguously interpreted as the presence of life. This is a drastic oversimplification, as false positives are expected to be pernicious obstacles that need to be overcome in order to assess the reliability of any feature. Common methods are to search for a degree of redundancy with correlated signatures that may boost our confidence that the signal we measure is biotic (see \cite{harman2018biosignature} for a recent review). But what is the desired degree of confidence when it comes to inferring the $f_i$s?  Obviously, the higher confidence the better, but in this section we show that there is a diminishing return beyond a certain point, and delineate the conditions for when effort should be expended in raising the confidence levels rather than optimizing some other part of the mission.
	
	To address this, we again restrict our attention to a single transition. As an example, this could be the advent of photosynthesis. As discussed above, the primary signature of this is detection of oxygen in a planet's atmosphere, but myriad abiotic processes are capable of oxygen production as well. Searching for correlated signatures such as water vapor and other gases would reduce this ambiguity but would be more difficult, require a more expensive mission, and would take more time which may potentially be used to explore other systems. Let us say, then, that for the measurement of some abstract biosignature, the confidence that can be ascribed to it being biotic in origin can be determined to be $c$. We also take the opportunity to define the diffidence $d=1-c$. In this case, if we detect $\ndet$ systems with this signature in our mission, we would only expect $\nbio\approx c\ndet$ to be biotic. Though this will hold on average, imperfect confidence introduces additional uncertainty in the number of systems which have attained the biogenic state we wish to measure. The main problem here is that we observe $\ndet$, but wish to infer $\nbio$, which determines $f$. To do this, we must average over each individual case, in which the sample we observe can have any number due to biotic and abiotic effects, weighted by the probability for each. The distribution for $f$ will be given by
	\beq
	p(f|\text{data},c)=\sum_{\nbio=0}^{\ndet}B(\ndet,\nbio,c)\,\beta(\ntot,\nbio,f)\label{poc}
	\eeq
	This is simply the marginalization over the unobservable number $\nbio$, which sums over all possible values that are consistent with the number of detections $\ndet$. This summation can be performed by explicitly using eqn (\ref{pB}) for each distribution, leading to the pdf being expressible in terms of hypergeometric functions:
	\beqa
	p(f|\text{data},c)=(\ntot+1)(1-c)^{\ndet}(1-f)^{\ntot}\,\times\nonumber\\\times\,{}_2\text{F}_1\left(-\ndet,-\ntot,1,\frac{c\,f}{(1-c)(1-f)}\right)\label{2f1}
	\eeqa
	It will not really be necessary to use this full form in this section, but it will be used in later sections.
	
	Then, the average can be determined from
	\beq
	\langle f\rangle=\sum_{\nbio=0}^{\ndet}B(\ndet,\nbio,c)\,\langle\, f\,|\,\nbio\,\rangle=\frac{c\,\ndet+1}{\ntot+2}
	\eeq
	where we have used eqn (\ref{avg}) for $\langle f|\nbio\rangle$, the average given $\nbio$ systems that have life. This indeed goes to $\langle f \rangle\rightarrow c\, \rdet$ in the large survey limit, in agreement with our expectations.
	
	A similar calculation may be carried out for the variance, yielding
	\beq
	\sigma_f^2=\frac{\langle f\rangle\,\Big(1-\langle f\rangle+d\,\Big)}{\ntot}
	\eeq
	where we have taken the large $\ntot$ limit. This reduces to the previous expression (\ref{avg}) when the confidence is exactly 1, but includes an additional contribution to the variance from the diffidence of our measurements. If the goal is to minimize this, $\langle f\rangle$ should be treated as a fixed number, but either $\ntot$ can be increased, as usual, or the diffidence can be decreased (confidence increased). This gives conditions for which should be preferred: since the diffidence contributes to the variance additively with respect to the usual $1-f$ term, the gain in precision from increasing the confidence only occurs while $c<\langle f\rangle$. Expending effort to increase confidence beyond this value will lead to diminishing returns. If one takes the ratio of maximally confident to maximally diffident standard deviations, for instance, one finds it to be $\sigma_f(d=0)/\sigma_f(d=1)=\sqrt{(1-\langle f\rangle)/(2-\langle f\rangle)}$: this yields a factor of 1.4 for $f\rightarrow0$, but can be arbitrarily small for $f\rightarrow1$, a consequence of the fact that the ordinary contribution to the variance vanishes in this limit. This suggests that we ought to invest more time in raising our confidence of measurements of those transitions which have a high probability of occurring, and not bother as much with the relatively rare transitions.
	
	\subsection{Confidence of the Confidence}
\begin{centering}
		\begin{figure*}
			\centering
			\includegraphics[width=.6\textwidth]{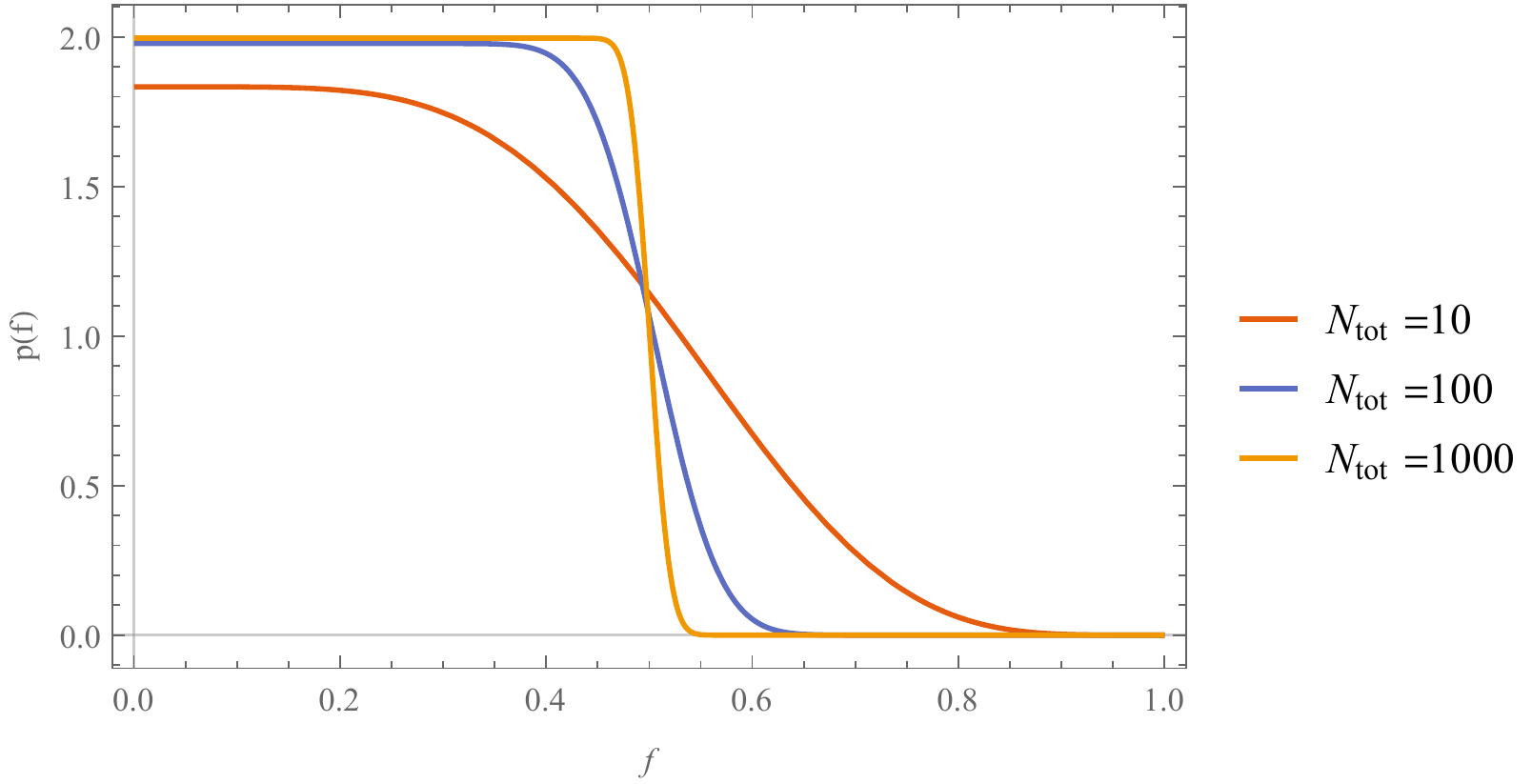}
			\caption{The inferred pdf of $\fbio$ as given by eqn (\ref{pocc}) when marginalizing over a uniform $p_c(c)$. Here $\rdet=1/2$. As can be seen, the full distribution approximately mimics the confidence prior, scaled to be between 0 and $\rdet$. This effect persists, and indeed becomes more exact, for large $\ntot$.}
			\label{noc}
		\end{figure*}
	\end{centering}

	Before, we were treating the confidence $c$ as if it were a known quantity with which we could assess the number of signals in our sample which were produced biotically. However, this is likely to not be the case. To account for this, we must marginalize the expression (\ref{poc}) over all possible values, weighted according to the probability we assign to any given value of the confidence. In this instance, we have
	\beq
	p(f|\text{data})=\int_0^1\text{d}c\,p_c(c)\sum_{\nbio=0}^{\ndet}B(\ndet,\nbio,c)\,\beta(\ntot,\nbio,f)\label{pocc}
	\eeq
	For instance, if the confidence is completely unknown, it can be treated as a uniform random variable $c\sim\mathcal U(0,1)$. This expression acts as an approximate step function, making any value of $f$ less than $\ndet/\nbio$ equally probable, and excluding values above this threshold, so that $p(f)\approx\mathcal U(0,\ndet/\ntot)$. This is plotted in Fig. \ref{noc}, where it can be seen that the width of the transition regime decreases with increasing $\ntot$. While a measurement like this would provide us with an upper bound on $\fbio$, this situation is far from ideal, and so it is recommended that at least some effort be spent on determining the value of $c$ before the measurement takes place.
	
	From the figure, we can see that this information erasure persists in the large sample limit, being a property of marginalizing over uncertainties in the setup rather than arising from measurement error. For a more general $p_c(c)$, the resultant distribution for $f$ will take the same form, only scaled to vanish for $f>\rdet$. This can be seen by noting that in the large sample limit, $B(\ndet,\nbio,c)\rightarrow\delta_{\nbio,c\ndet}$ and $\beta(\ntot,\nbio,f)\rightarrow\delta(\nbio-f\ntot)$, so that eqn (\ref{pocc}) gives $p(f|\text{data})\rightarrow p_c(f/\rdet)/\rdet$. From this we conclude that in order to maximize information gain we would want to make $p_c(c)$ as close to a delta function as possible.
	
	\subsection{Conditional confidence}
	
	A further complication is introduced when the confidence depends on the fraction of systems which possess life, which is unknown before the measurement is made. In this case, the confidence that our signal is biotic is related to $\fbio$, the quantity we are trying to measure!  This can be summed up in the expression
	\beq
	c=\frac{\fbio}{\fbio+\fabio}\label{candfbio}
	\eeq
	where $\fbio$ is the fraction of systems that produce the signal in question biotically, and $\fabio$ is the fraction of systems that produce the signal abiotically. This was discussed in \cite{catling2018exoplanet} and \cite{walker2018exoplanet}, who use a Bayesian framework for inferring $\fbio$. From here, we can see that if enough care is taken to remove all false positives, by searching for enough independent lines of evidence that we can be certain the signal could not have been produced abiotically, then $c\rightarrow1$. Apart from that, then our confidence in the signal depends on the value $\fbio$ itself. This subtlety is actually rather straightforward to deal with: one may simply substitute this expression for $c$ into the probability density for $f$ given by eqn (\ref{2f1}) ( taking care to set the normalization of the distribution so that it integrates to 1). However, when this is done, an alarming conclusion is reached: the distribution actually peaks at two places: one at $f\sim \rdet= \ndet/\ntot$, and the other at $f=0$!  This effect persists even for large $\ntot$, and can be understood as follows: imagine the simplified scenario where we have detected $\ndet$ stars from a sample of $\ntot$, but are unable to determine between the case $c=0$ and $c=1$. Then, our inferred value of $\fbio$ will be given by
	\beq
	p(\fbio|\ndet)=\alpha {\ntot\choose\ndet}f^{\ndet}(1-f)^{\ntot-\ndet}+(1-\alpha)(1-f)^{\ntot}
	\eeq
	where $\alpha$ is our confidence that $c=1$. Even in this simple setting, two peaks are evident, at $\fbio=\rdet-\fabio$ and $\fbio=0$. Furthermore, since the second peak is a stronger function, it can carry more weight than the other unless $\alpha$ is close enough to 1. This tells us that unless we are sure enough in our systematic account for the signal's origin, the conclusion of our measurement is that it almost certainly occurred abiotically. In this situation, it would not be necessary to invoke the presence of life to explain any of the detections we make.
	
	\begin{centering}
		\begin{figure*}
			\centering
			\includegraphics[width=.6\textwidth]{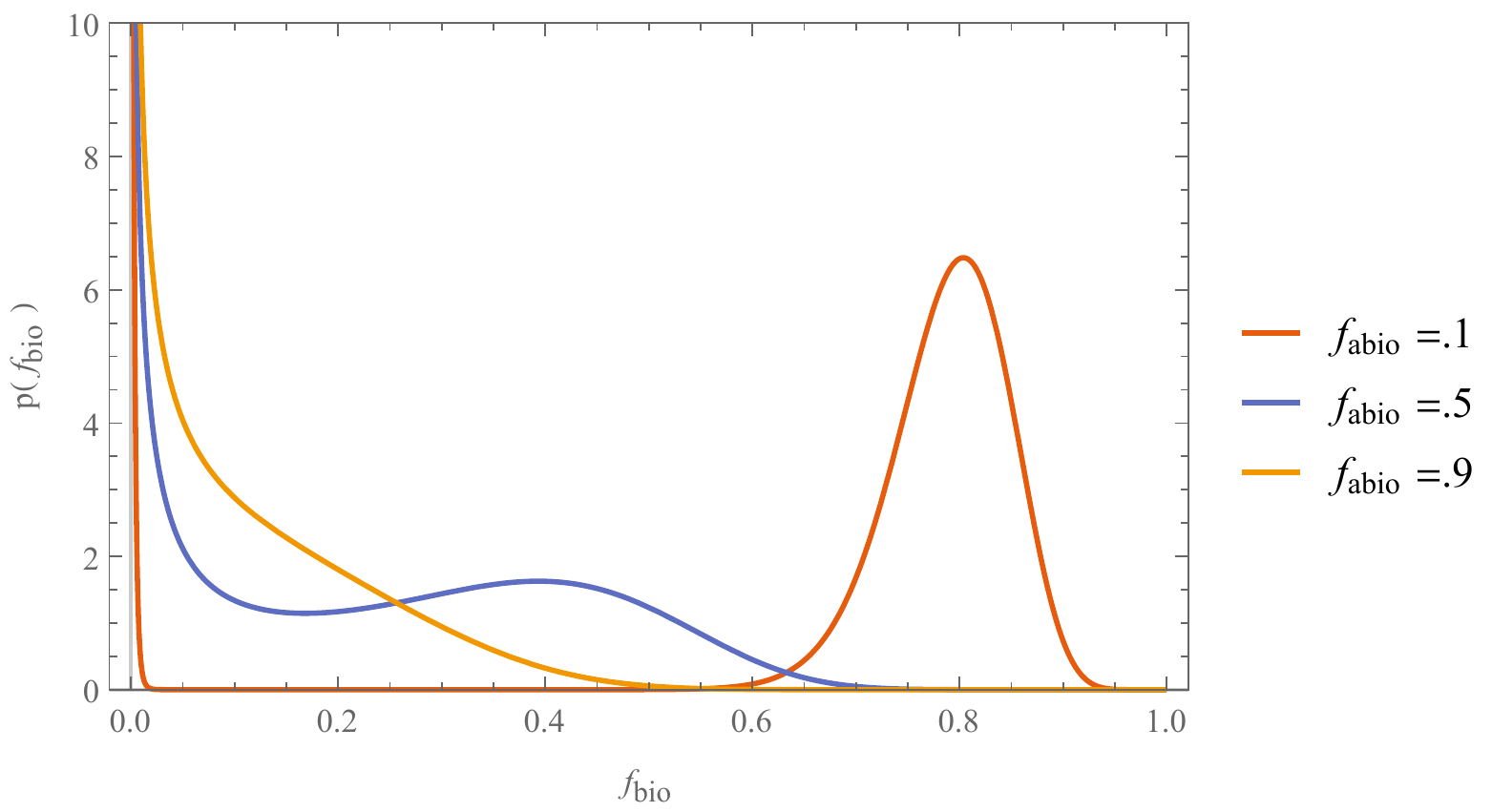}
			\caption{The pdf of $\fbio$ when the confidence is related to $\fbio$,  found by inserting expression (\ref{candfbio}) into the distribution (\ref{2f1}). All graphs here have $\ntot=100$ and $\rdet=.9$. Notice that the peak at 0 still persists, even for the $\fabio=.1$ case, and that the main peak is offset from the observed by the fraction of abiotic signals. The desired situation is when these two peaks are well separated and most of the weight is in the one centered on the nonzero value, which occurs when $\fabio\ll\rdet$.}
			\label{coff}
		\end{figure*}
	\end{centering}
	
	In order to quantify the certainty required to avoid this situation, let us return to the full analysis: several example distributions are depicted in Fig. \ref{coff}.
	
	In order to estimate the relative likelihood that $\fbio\approx0$, we can make a crude linear approximation around $\fbio=0$ to determine the total probability that the inferred fraction is within this regime:
	\beq
	p(\fbio|\text{data})\approx(\ntot+1)\left(1-\left(\ntot+\frac{\ndet}{\fabio}\right)\fbio\right)+\mathcal{O}(\fbio^2)
	\eeq
	which gives
	\begin{equation}
	p(\fbio\sim0|\text{data})\approx\frac12\frac{\ntot+1}{\ntot+\frac{\ndet}{\fabio}}\rightarrow \left\{
	\begin{array}{cl}
	\frac12 & \fabio\gg\rdet\\
	\frac{\fabio}{\rdet} & \fabio\ll\rdet
	\end{array} \right.
	\end{equation}
	From here, we can see that in order to avoid any contribution from the $\fbio\sim0$ peak, we would like to be in the $\fabio\ll\rdet$ regime. This is intuitively clear, as since the peaks of the pdf occur at 0 and $\fbio\approx\rdet-\fabio$, they will merge if the fraction of abiotic signals is close to the detected fraction into a single peak at $\fbio\approx0$.
	
	\subsection{False Negatives}
	
	Just as there may be false positive biosignatures, there may be false negatives as well. These can come about if an existing biosphere does not produce the signal we expect of it, in which case it is termed cryptic (\cite{cockell2014habitable}), or if the signal is produced but obscured, say by clouds or haze (\cite{zugger2011searching}). It is worthwhile to bear in mind that due to haze, many of the proposed biosignatures would have been unobservable on Earth during the Proterozoic eon (\cite{reinhard2016earth}), which represented a substantial fraction of Earth's history. For this reason it will be important to understand obscuration as well as we possibly can, and efforts to understand the diversity of biological and planetary environments will mitigate some of these uncertainties. However, even if these factors are understood perfectly, there will still be considerable uncertainty in the observed number of exoplanet biospheres: if a seemingly lifeless planet is mostly obscured, how sure can we be that it actually does not possess life?  We outline the statistical framework for how to handle this here.
	
	If we survey $\ntot$ planets and make $\ndet$ detections of life, this leaves $\ntot-\ndet$ where no life has been detected. If our confidence that life is absent on each of these is given by $c_a$, then the number of inhabited planets in our sample is between $\ndet\leq\nbio\leq\ntot$. The distribution for the fraction of inhabited planets is given by
	\beqa
	p(\fbio|\text{data},c_a)=\sum_{\nbio=\ndet}^{\ntot}B(\ntot-\ndet,\ntot-\nbio,c_a)\,\times\nonumber\\\times\,\beta(\ntot,\nbio,\fbio)\label{pou}
	\eeqa
	This resembles the result we had for false positives in eqn (\ref{poc}), and reduces to the standard beta distribution in the limit that $c_a\rightarrow1$.
	
	The average and variance can be computed for this distribution: the average is
	\beq
	\langle\fbio\rangle=\rdet+(1-c_a)(1-\rdet)
	\eeq
	and the variance
	\beq
	\sigma^2\rightarrow\frac{(\langle\fbio\rangle+d_a)(1-\langle\fbio\rangle)}{\ntot}
	\eeq
	If we wish to minimize the uncertainty of our measurement, we should minimize the diffidence $d_a=1-c_a$, but only to the point where it is the same order as $\langle\fbio\rangle$. Notice a key difference in interpretation here: if $c_a\rightarrow0$, $\langle\fbio\rangle\rightarrow1$, independent of the number of biosignatures measured. In this limit, our complete lack of confidence implies that we are actually certain that the lack of signal from a given planet is due to life being present there, but the signal being obscured. This is a somewhat pathological limit, and does not express ignorance as well as the $c\rightarrow1/2$ limit, for example.
	
\begin{centering}
		\begin{figure*}
			\centering
			\includegraphics[width=.6\textwidth]{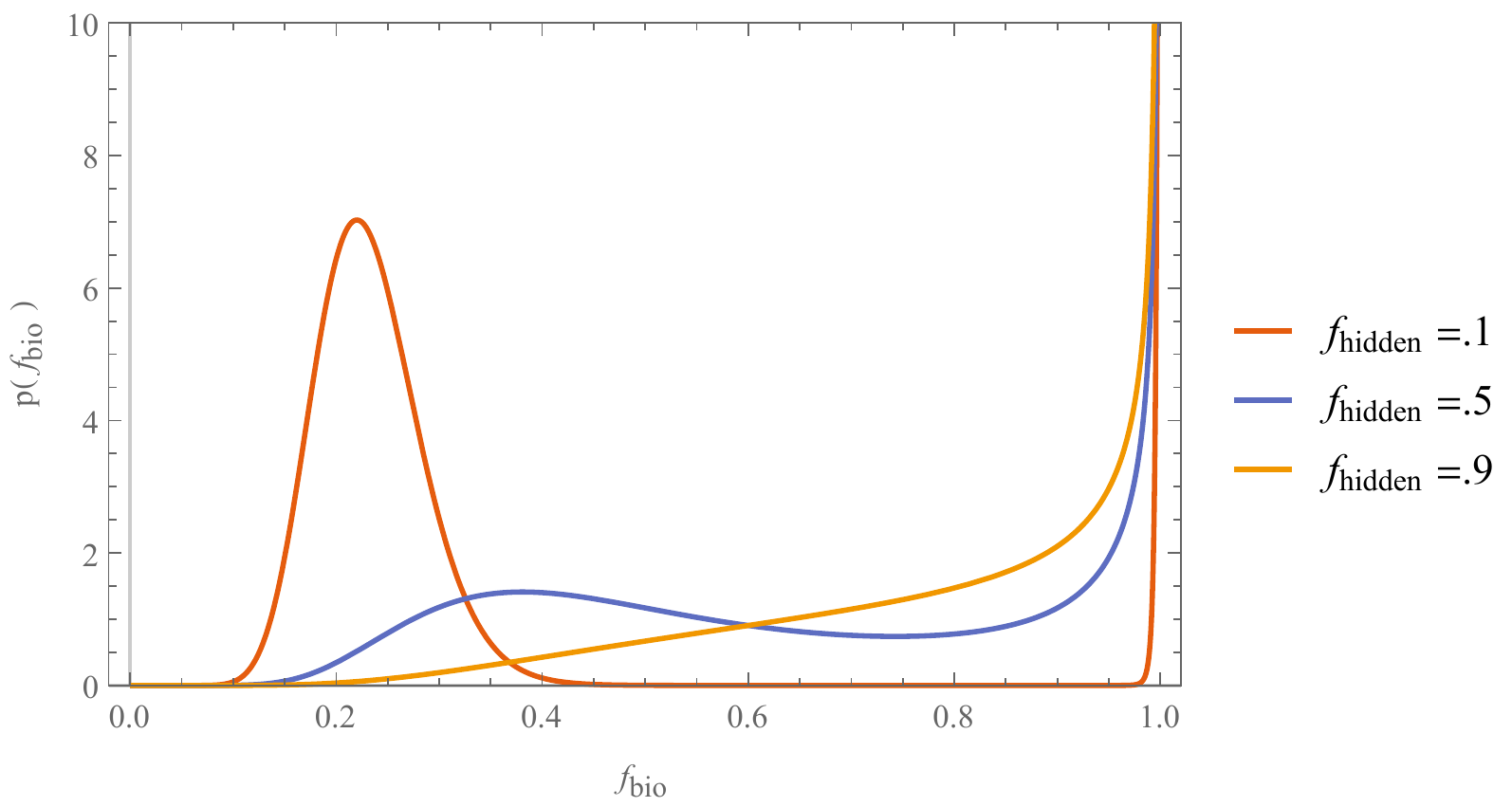}
			\caption{Probability distributions for $\fbio$ with false negatives of various degrees of confidence taken into account, as found by inserting eqn (\ref{ca}) into the distribution (\ref{pou}). Here, $\ntot=100$ and $\ndet=20$ throughout. As with the false positive case, peak around the detected value dominates only when $f_\text{hidden}\ll\fbio$.}
			\label{hide}
		\end{figure*}
	\end{centering}

	As before, the confidence in the interpretation of the absence of a biosignature depends on the likelihood of life occurring, which is the quantity we set out to measure. If the confidence is given by 
	\beq
	c_a=\frac{f_\text{absent}}{f_\text{absent}+f_\text{hidden}\fbio}=\frac{1-\fbio}{1-(1-f_\text{hidden})\fbio}\label{ca}
	\eeq
	then the distribution for $\fbio$ can be found with this substitution into eqn (\ref{pou}) (again, by altering the normalization to ensure that it integrates to 1). This is plotted for several values of $f_\text{absent}$ in Fig. \ref{hide}. Here, a peak at $\fbio\approx1$ is prominent unless $f_\text{hidden}<\fbio$. The explanation is the same as for the false positive case.
	
	\subsection{Total Number Unknown}
	
	Another scenario that may occur is that the number of detections could be precisely measured, yet the total number in the sample may be unknown. An example of this would be if the mass of the planet were determined through radial velocity techniques only in the combination $m\,\text{sin}\,i$, with $i$ the inclination angle, it would not be clear whether the planets in the sample are Earthlike or not. Additionally, any false positives or negatives in surveys of primitive biosignatures would manifest as an unknown total number in surveys of more advanced ones. Here, we outline the effects of this type of uncertainty, and how best to mitigate it.
	
	For our setup, we will take $\ndet$ as the number of detections, $c_T$ as the confidence that a nondetection should be included in the total count, and $\ntop$ as the upper limit of possible systems in the sample. The inferred distribution for $\fbio$ will then be
	\beqa
	p(\fbio|\text{data},c_T)\,\propto\sum_{\ntot=\ndet}^{\ntop}B(\ntop,\ntot,c_T)\,B(\ntot,\ndet,f)\nonumber\\=B(\ntop,\ndet,c_T\,f)
	\eeqa
	Here, we have taken the lower limit to be the number of detections in the sample: in general, this could be an arbitrary number instead, but the analysis is complicated considerably, and the insights we glean from this simpler exercise hold in the more general case. Now, the average can be given in terms of incomplete beta functions:
	\begin{equation}
	\langle\fbio\rangle=\frac{\beta_{c_T}(\ndet+2,\ntop-\ndet+1)}{c_T\,\beta_{c_T}(\ndet+1,\ntop-\ndet+1)}\rightarrow \left\{
	\begin{array}{cl}
	\frac{\rdet}{c_T} & c_T\rightarrow 1\\
	1 & c_T\rightarrow 0
	\end{array} \right.
	\end{equation}
	Where $\rdet=\ndet/\ntop$. These limits are as expected. The variance is similarly given by
	\begin{eqnarray}
	\sigma_{\fbio}^2=\frac{\beta_{c_T}(\ndet+3,\ntop-\ndet+1)}{c_T^2\,\beta_{c_T}(\ndet+1,\ntop-\ndet+1)}-\langle\fbio\rangle^2\nonumber\\\rightarrow \left\{
	\begin{array}{cl}
	\frac{\rdet(1-\rdet)}{c_T^2\,\ntop} & c_T\rightarrow 1\\
	\frac{1}{\ntop^2} & c_T\rightarrow 0
	\end{array} \right.
	\end{eqnarray}
	This latter limit is somewhat unusual, since here the total number in the sample is bound to be $\ndet$. The real question, though, is what value of $c_T$ sets the transition between these two different behaviors. This can be determined by finding the subleading corrections to the asymptotic expressions above. When this is done, it is found that these expressions lose their validity when $c_T\approx\rdet$. As before, the desired confidence, even in this slightly different setting, should be greater than the observed ratio in order to garner useful information from the measurement.

	\section{Discussion}\label{disc}
	
	Let us summarize the lessons we've garnered from sections \ref{sing}-\ref{conf}.  We first noted, somewhat obviously, that uncertainty in measurement of any biosignature decreases with sample size.  An observation with somewhat more content is that when measuring multiple biosignatures, if a bottleneck transition is present, then the uncertainties in all the transitions that occur after that are doomed to be large, and so not much will be gained in trying to measure them.  When combining datasets that both measure two biosignatures, the variances are proportional to the combined survey sizes.  If one survey only measures the more primitive biosignature, then the uncertainty in that depends on a sum of the two survey sizes, while the second biosignature depends only on the size of the second survey. If the first survey only measures the more primitive and the second the more advanced biosignature, then uncertainty in the more advanced is a sum of two terms, and minimizing this will usually occur when the two are equal.  When false positives are taken into account, the measurement uncertainty picks up an additional term that becomes important when the uncertainty in interpretation is of order the signal prevalence. This conclusion holds true when considering false negatives, false samples and signal confidence which is conditional on the rate of occurrence: it behooves a survey planner to ensure that the confidence in interpretation is greater than the measurement, but not necessarily much more than that, if resources could instead be devoted to increasing the survey volume.  These general lessons will hopefully bear useful to bear in mind when future missions are being designed.
	
	The stated goal of many missions is to ``maximize science return''. When attempting something that has never been done before, merely accomplishing the goal once, in the cheapest and most guaranteed manner possible, suffices. Once this is done, the method can be extended to larger data sets, with the ultimate goal of generating new knowledge. In the context of measuring atmospheres of Earthlike exoplanets, we are still at the stage where satisfying the task should be our top priority. With the next generation of experiments, it will be possible to obtain the spectra of dozens of Earthlike worlds, but we should not be too discriminate about which are worthy of observation time: any within reach should be targeted. Moving beyond that, when the characterization of hundreds of worlds is possible, we should rather focus on studying those that we suspect can teach us something qualitatively new.
	
	Is it possible to formalize this intuition?  This is the task we are trying to accomplish above. Our main point is that data about various biosignatures can be used to infer values for the fraction of planets undergoing successive biosphere transitions. The uncertainties in these measurements serve as natural candidates for the quantities to be minimized.
	
	Though we have illustrated how this approach may be used to given design and/or target selection recommendations in several idealized scenarios, our discussion to this point has made no pretense at providing a comprehensive framework. In particular, since there are actually multiple simultaneous quantities to be measured, which one should be prioritized?  Here, we offer a proposal: the science return of a mission can, loosely, be equated with the information that is gained. We propose that this sentiment should be taken literally, in the technical sense. Then the science return can be defined as $S=-\int \text d^nf \,p(f_i|\text{data})\log p(f_i|\text{data})$, where $p$ is the joint pdf of all measured quantities. Though this is somewhat cumbersome to compute, an analytic expression for science return can be found. In the case of a single biosignature with confidence $c=1$, eqn (\ref{pbeta}) can be used to compute $S$ as:
	\beqa
	S=\log\left(\frac{(\ntot+1)!}{\ndet!(\ntot-\ndet)!}\right)+\ndet\, H(\ndet)\nonumber\\-\ntot\, H(\ntot+1)+(\ntot-\ndet) \,H(\ntot-\ndet)\label{Seq}
	\eeqa
	where $H(n)=\sum_{k=1}^n1/k$ is the nth harmonic number. In the large survey limit, the full science return can be approximated as
	\beq
	S\approx -\frac12\sum_i\log\sigma_i^2
	\eeq
	where the variance of each quantity is summed. This definition has a number of attractive features. Firstly, in the standard case of an unambiguous measurement of a single quantity, the variance is given by that of a binomial distribution, eqn (\ref{avg}), and maximizing the science return corresponds simply to increasing the number of samples as much as possible. When there are multiple measured quantities, if we can abstractly parameterize how each depends on a single quantity $e$, denoting effort, then, if the derivative of each is denoted by $R_i=d\sigma_i/de$, we have $dS/de=-\sum_i R_i/\sigma_i$. To maximize, effort should be invested into whichever term in this sum is the largest, corresponding to whichever has a large $R_i$ or a small $\sigma_i$: where improvement is easy and precision is possible.
	
	In Table \ref{Science}, we display the science return for the upcoming missions we referred to in section \ref{tele}. This table treats biosignature detection as having absolute certainty, but folding additional analyses such as confidence levels into this expression would merely shift the numbers, leaving the overall trends unaltered. Note that in the first half of the table, the information gain is greater for smaller $f_i$; this is because, when measuring the rate of a rare a event and a common event to the same precision, the measurement of the rare event yields more information. In the lower half of the table, however, more information is gained if the fractions are more common; this is because these missions will measure multiple biosignatures, which will probably be nonexistent in the samples if the fractions are small. This should give some indication of the merit of these missions, though again we stress that these numbers are only in terms of the particular question we pose, and are meant as a rough heuristic only. Taking this quantity too seriously can run afoul of Goodhart's law of perverse incentives, but as a heuristic it can serve to quickly clarify which features of a complex mission should be improved.
	
	\begin{table}
		\vskip.4cm
		\begin{center}
			\begin{tabular}{|c|c|c|}
				\hline 
				telescope & $S(f_i=1/2)$ & $S(f_i=1/10)$ \\
				\hline
				JWST & .18 & .44\\
				WFIRST & .39 & .80\\
				GMT & .47 & .93\\
				TMT & .47 & .93\\
				E-ELT & .81 & 1.39\\
				HabEx & 1.47 & 1.76 \\
				OST (6 m)& 1.29 & 1.59 \\
				OST (9 m)& 2.03 & 2.31 \\
				LUVOIR (8 m) & 3.33 & 3.61 \\
				LUVOIR (15 m) & 4.22 & 4.51 \\
				RAVE/GAIA & 4.97 & 4.54 \\
				ELF (20 m) & .91 & 1.51 \\
				ELF (70 m) & 4.11 & 4.40 \\
				OWL-MOON & 7.38 & 7.69 \\
				Hypertelescope & 13.02 & 13.33 \\
				FOCAL & 19.34 & 19.65 \\
				\hline
			\end{tabular}
		\end{center}
		\caption{The expected science return for different proposed telescope technologies, as given by eqn (\ref{Seq}). Different columns correspond to different values for the actual fractions of each type of biosignature, and we have used log base 2 so as to assign an interpretation as the number of bits of information that a given mission will return.}
		\label{Science}
	\end{table}
	
	Though our discussion was highly idealized, we were capable of gleaning several lessons. We were able to quantify how large our sample should be to avoid prior bias, which signals will be too weak to bother measuring, which survey to spend time maximizing when combining two, and how well we should establish the confidence of our measurements. Though our recommendations are always phrased in terms of desired relative sample sizes or degrees of confidence, if this is coupled with a model for how these scale with price for a particular telescope, more concrete recommendations can be made. Extensions to our work could be equally profitable. Perhaps the most urgent task is to relax the assumption that all exoplanet systems are identical, and allow the fractions we discuss to depend on environmental variables. Extending our analysis to cover this scenario should be capable of yielding recommendations on which hypothesized trends to focus on measuring in upcoming datasets. \\
	
	{\bf \noindent Acknowledgements}
	We would like to thank Jean-Philippe Beaulieu, Ashley Baker, Cullen Blake, Marc Postman, and Ileana P\'erez-Rodr\'iguez for helpful discussions. MS is grateful to the Balzan Centre for Cosmological Studies and the Institut d’Astrophysique de Paris for their support during the completion of this work.

	
	
	
	\bibliographystyle{mnras}
	\bibliography{search} 


	\bsp	
	\label{lastpage}
\end{document}